\documentclass[aps,pre,twocolumn,superscriptaddress,floatfix]{revtex4-1}
\usepackage[T1]{fontenc}
\usepackage[latin9]{inputenc}
\usepackage{fancyhdr}
\usepackage{float}
\usepackage{amsmath}
\usepackage{cancel}
\usepackage{graphicx}



\begin{document}
\title{Localization in finite asymmetric vibro-impact chains}
\author{Itay Grinberg}
\email{GrinbergItay@gmail.com}
\author{Oleg V. Gendelman}
\email{ovgend@technion.ac.il}
\affiliation{Faculty of Mechanical Engineering\\ Technion - Israel Institute of Technology}
\date{\today}
\begin{abstract}
We explore the dynamics of strongly localized periodic solutions (discrete solitons, or discrete breathers) in a finite one-dimensional chain of asymmetric vibro-impact oscillators. The model involves a parabolic on-site potential with asymmetric rigid constraints (the displacement domain of each particle is finite), and a linear nearest-neighbor coupling. When the particle approaches the constraint, it undergoes an impact (not necessarily elastic), that satisfies Newton impact law. Nonlinearity of the system stems from the impacts; their possible non-elasticity is the sole source of damping in the system. We demonstrate that this vibro-impact model allows derivation of exact analytic solutions for the asymmetric discrete breathers, both in conservative and forced-damped settings. The asymmetry makes two types of breathers possible: breathers that impact both or only one constraint. Transition between these two types of the breathers corresponds to a grazing bifurcation. Special character of the nonlinearity permits explicit derivation of a monodromy matrix. Therefore, the stability of the obtained breather solutions can be exactly studied in the framework of simple methods of linear algebra, and with rather moderate computational efforts. All three generic scenarios of the loss of stability (pitchfork, Neimark-Sacker and period doubling bifurcations) are observed.
\begin{description} \item [{PACS~numbers}] 05.45.Yv, 63.20.Pw, 63.20.Ry{\small \par} \end{description} \end{abstract}
\maketitle

\section*{Introduction}

Localization is an important and widely studied phenomenon in discrete
dynamical systems \cite{Flach1998,Flach2008,Vakakis1996,Campbell2004,Vakakis2008,Anderson1958,PIERRE1987549,Bendiksen2000,Cai1994}.
Contrary to the localization in linear systems, in nonlinear systems
it is possible without any disorder, i.e. localization may occur even
in a purely homogeneous nonlinear lattice. Interesting examples of
such localized responses are Discrete Breathers (DBs), sometimes referred
to as Intrinsic Localized Modes (ILMs) or Discrete Solitons. The DB
is a periodic strongly localized response of the lattice system. It
can be simply imagined as an oscillating envelope localized in the
vicinity of a single or several sites of the lattice. The DB's localization
is typically exponential; however, in the systems with strong nonlinearity,
it may be hyper-exponential\cite{Flach2008}. The DBs are known in
various branches in physics. They were experimentally observed and
theoretically discussed in a variety of model systems, such as superconducting
Josephson junctions \cite{Trias2000}, nonlinear magnetic metamaterials\cite{Lazarides2006},
electrical lattices\cite{English2012}, micro-mechanical cantilever
arrays\cite{Gutschmidt2010,Kimura2009a,Kenig2009,Sato2011,Sato2006},
Bose-Einstein condensates\cite{Trombettoni2001}, and chains of mechanical
oscillators\cite{Gendelman2008,Cuevas2009,Gendelman2013,Grinberg2016,Perchikov2015}.

The exact DB solutions in specific nonlinear chain models remain scarce
due to the nonlinearity and discreteness of the systems that encumbers
the derivation of exact solutions. Current theoretical research primarily
concentrates on numerical explorations and approximate analytic approaches
\cite{Flach1998,Flach2008,Cai1994,Romeo2016}. Few known exceptions
are the completely integrable Ablowitz-Ladic model\cite{Ablowitz1976},
chains with homogeneous interaction\cite{Ovchinnikov1999}, and vibro-impact
chains\cite{Gendelman2008}. Recently the latter approach was extended
to the forced-damped vibro-impact chains\cite{Gendelman2013,Perchikov2015},
chains with self-excitation\cite{Shiroki2016}, and, most recently,
to Multi-Breather (MB) solutions \cite{Grinberg2016}, namely the
DBs with more than a single localization site.

The aforementioned vibro-impact chains are essentially linear, except
for the possibility of collisions, i.e. all the on-site and coupling
interactions, that are not impacts, are linear. This feature not only
allows the derivation of the exact solution, but also considerably
simplifies the stability analysis. The stability of the periodic DB
solution is determined by location of the eigenvalues of the Monodromy
matrix \cite{Strogatz1994}. In most cases, the monodromy matrix can
only by obtained numerically by integration of the equations of motion
over the period of the solution. This task can be extremely difficult
when treating systems with a large number of particles, to the extent
that super-computers may be necessary. The considered vibro-impact
models allow explicit derivation of the monodromy matrix \cite{Gendelman2013,Perchikov2015,Grinberg2016}.
Thus the stability analysis is reduced to evaluation of the spectrum
of easily computed matrices. Consequently, even simple PCs suffice
for chains with thousands of particles. Furthermore, even for
the finite chains, one can accomplish these derivations without further
approximations.

This work is based on the approach used in refs. \cite{Gendelman2008,Gendelman2013,Perchikov2015,Grinberg2016},
but introduces important novel feature into the model. In symmetric
models, a breakdown of the DB symmetry is one of the instability scenarios
\cite{Gendelman2013}. In current model, the asymmetry is imbedded
into the lattice itself. To be more specific, the model comprises
a finite number of linearly coupled oscillators; each of the latter
includes on-site coupling with two asymmetric rigid barriers bounding
the movement. Thus one can obtain the asymmetric DBs and explore their
zones of existence and stability properties in the space of parameters.
Furthermore, the considered asymmetry allows a new type of the DB
\textendash{} the single-sided DB \textendash{} where the impact only
occurs at one of the constraints. In sec. \ref{sec:Description} we
present a detailed description of the system. For this new type of DBs we observe for the first time the period doubling bifurcations and obtain an analytic solution for the emerging solution in a similar manner. Section \ref{sec:Exact-Solution}
contains the derivations of an exact analytic solution for both conservative
and forced/damped DBs, as well as for the single-sided DB. The method
of the stability analysis is explained in sec. \ref{sec:Stability}.
Numerical validation of the results is presented in sec. \ref{sec:Numerical-Validation},
followed by concluding remarks in sec. \ref{sec:Concluding-Remarks}.

\section{\label{sec:Description}Description of the Model}

We consider a chain of $\left(N+1\right)$ identical unit masses,
coupled to their neighbors via linear springs, and with periodic boundary
conditions. In addition, all masses are subject to identical on-site
potentials. The on-site interaction is via linear spring, but the
motion is bounded by a set of asymmetric impact barriers. The on-site
and coupling potentials can be described as follows:
\begin{equation}
V{\left(u\right)}=\begin{cases}
\gamma_{1}u^{2} & \,\,\,\,\,\left|u-a\right|<1\\
\mbox{Impact} & \left|x-a\right|=1
\end{cases}
\end{equation}
\begin{equation}
W{\left(u\right)}=\gamma_{2}u^{2}
\end{equation}
where $a\geq0$ is the parameter of asymmetry and $\gamma_{1}$ and
$\gamma_{2}$ are the on-site and coupling stiffnesses, respectively.

This yields the following Hamiltonian for the full finite chain:
\begin{equation}
\begin{array}{c}
H=\underset{n=0}{\overset{N}{\sum}}{\left(\cfrac{1}{2}p_{n}^{2}+V{\left(u_{n}\right)}\right)}+\\
+\underset{n=0}{\overset{N-1}{\sum}}{W{\left(u_{n}-u_{n+1}\right)}}+W{\left(u_{N}-u_{0}\right)}
\end{array}\label{eq:3}
\end{equation}

The impact, which could be either elastic or non-elastic, obeys the
following traditional Newton impact law:
\begin{equation}
\dot{u}{\left(t_{i}+\right)}=-e\dot{u}{\left(t_{i}-\right)}
\end{equation}
where $t_{i}$ is the time instance of the impact and $0<e\leq1$
is the coefficient of restitution.

\section{\label{sec:Exact-Solution}Exact Solution for the Asymmetric Breather}

\subsection{Discrete Breathers in the Conservative Model}

If the external forcing is absent, one should set the coefficient
of restitution to $e=1$ in order to preclude the dissipation. Equations
of motion for the chain can be easily obtained from the Hamiltonian
given in \eqref{eq:3}. Furthermore, the periodicity of the DB allows
introducing the impact into the equations of motion as external forcing
in the the form of a sum of delta functions with advanced-delayed
arguments. Without restricting the generality, we adopt that the DB
is localized at the particle with $n=0$. One obtains the following
equations of motion:
\begin{equation}
\begin{array}{c}
\ddot{u}_{0}+\gamma_{1}u_{0}+\gamma_{2}\left(2u_{0}-u_{1}-u_{N}\right)=\\
=2p_{1}\sum_{j=-\infty}^{\infty}{\delta{\left(t-\phi-\cfrac{2\pi j}{\omega}\right)}}-\\
-2p_{2}\sum_{j=-\infty}^{\infty}{\delta{\left(t-\cfrac{2\pi j}{\omega}\right)}}
\end{array}
\end{equation}
\begin{equation}
\ddot{u}_{n}+\gamma_{1}u_{n}+\gamma_{2}\left(2u_{n}-u_{n+1}-u_{n-1}\right)=0
\end{equation}
\begin{equation}
\ddot{u}_{N}+\gamma_{1}u_{N}+\gamma_{2}\left(2u_{N}-u_{0}-u_{N-1}\right)=0
\end{equation}

where $2p_{1}$ and $2p_{2}$ correspond to the amounts of momentum
transferred in the course of each of the impacts, $\delta$ is the
Dirac delta and $\phi$ is the phase instance of the secondary impact
in the period of the DB.

The expression for the impact forcing can be re-written in the form
of generalized Fourier series:
\begin{equation}
\begin{array}{c}
\ddot{u}_{0}+\gamma_{1}u_{0}+\gamma_{2}\left(2u_{0}-u_{1}-u_{N}\right)=\\
=\frac{\omega}{\pi}\sum_{j=-\infty}^{\infty}{\left(p_{1}\cos{\left(j\omega\left(t-\phi\right)\right)}-p_{2}\cos{\left(j\omega t\right)}\right)}
\end{array}
\end{equation}
\begin{equation}
\ddot{u}_{n}+\gamma_{1}u_{n}+\gamma_{2}\left(2u_{n}-u_{n+1}-u_{n-1}\right)=0
\end{equation}
\begin{equation}
\ddot{u}_{N}+\gamma_{1}u_{N}+\gamma_{2}\left(2u_{N}-u_{0}-u_{N-1}\right)=0
\end{equation}

Thus, one further obtains:
\begin{equation}
\begin{array}{c}
\ddot{u}_{0}+\gamma_{1}u_{0}+\gamma_{2}\left(2u_{0}-u_{1}-u_{N}\right)=\frac{\omega}{\pi}\left(p_{1}-p_{2}\right)+\\
+\frac{2\omega}{\pi}\sum_{j=1}^{\infty}{\left(p_{1}\cos{\left(j\omega\left(t-\phi\right)\right)}-p_{2}\cos{\left(j\omega t\right)}\right)}
\end{array}\label{eq:7}
\end{equation}
\begin{equation}
\ddot{u}_{n}+\gamma_{1}u_{n}+\gamma_{2}\left(2u_{n}-u_{n+1}-u_{n-1}\right)=0
\end{equation}
\begin{equation}
\ddot{u}_{N}+\gamma_{1}u_{N}+\gamma_{2}\left(2u_{N}-u_{0}-u_{N-1}\right)=0\label{eq:9}
\end{equation}

To obtain the exact solution, the displacement of each particle is
also searched in the form of Fourier series, and the following anzats
is used:
\begin{equation}
u_{n}=u_{n,0}+\sum_{j=1}^{\infty}{\left(u_{n,j,1}\cos{\left(j\omega\left(t-\phi\right)\right)}+u_{n,j,2}\cos{\left(j\omega t\right)}\right)}
\end{equation}
where,
\begin{eqnarray}
u_{n,0} & = & A_{0}f_{0}^{n}+B_{0}f_{0}^{-n}\\
u_{n,j,1} & = & A_{j}f_{j}^{n}+B_{j}f_{j}^{-n}\\
u_{n,j,2} & = & C_{j}f_{j}^{n}+D_{j}f_{j}^{-n}
\end{eqnarray}
Linearity of the equations of motion between the impacts yields:
\begin{equation}
\begin{array}{c}
f_{j}=\cfrac{\gamma_{1}+2\gamma_{2}-j^{2}\omega^{2}\pm\sqrt{\left(j^{2}\omega^{2}-\gamma_{1}-2\gamma_{2}\right)^{2}-4\gamma_{2}^{2}}}{2\gamma_{2}}=\\
=\cfrac{\gamma_{1}+2\gamma_{2}-j^{2}\omega^{2}\pm\sqrt{\left(j^{2}\omega^{2}-\gamma_{1}-4\gamma_{2}\right)\left(j^{2}\omega^{2}-\gamma_{1}\right)}}{2\gamma_{2}}
\end{array}
\end{equation}

Account of the periodic boundary conditions in eq. (\ref{eq:9}) yields
the following relations:
\begin{eqnarray}
A_{j} & = & B_{j}f_{j}^{-N-1}\\
C_{j} & = & D_{j}f_{j}^{-N-1}
\end{eqnarray}

Finally, substituting these derivations into the first equation of
system (\ref{eq:7}), that describes the dynamics of impacting mass,
one obtains for $j>0$:
\begin{eqnarray}
B_{j} & = & -\frac{2\omega p_{1}}{\pi\gamma_{2}\left(f_{j}-f_{j}^{-1}\right)\left(f_{j}^{-N-1}-1\right)}\\
D_{j} & = & \frac{2\omega p_{2}}{\pi\gamma_{2}\left(f_{j}-f_{j}^{-1}\right)\left(f_{j}^{-N-1}-1\right)}
\end{eqnarray}
and for $j=0$:
\begin{equation}
B_{0}=\frac{\omega\left(p_{2}-p_{1}\right)}{\pi\gamma_{2}\left(f_{0}-f_{0}^{-1}\right)\left(f_{0}^{-N-1}-1\right)}
\end{equation}

Summarizing, one obtains the following exact solution for the DB:
\begin{equation}
u_{n}=u_{n,0}+\sum_{j=1}^{\infty}{\left(u_{n,j,1}\cos{\left(j\omega\left(t-\phi\right)\right)}+u_{n,j,2}\cos{\left(j\omega t\right)}\right)}
\end{equation}
where
\begin{eqnarray}
u_{n,0} & = & \frac{\omega\left(p_{2}-p_{1}\right)\left(f_{0}^{n-N-1}+f_{0}^{-n}\right)}{\pi\gamma_{2}\left(f_{0}-f_{0}^{-1}\right)\left(f_{0}^{-N-1}-1\right)}\\
u_{n,j,1} & = & -\frac{2\omega p_{1}\left(f_{j}^{n-N-1}+f_{j}^{-n}\right)}{\pi\gamma_{2}\left(f_{j}-f_{j}^{-1}\right)\left(f_{j}^{-N-1}-1\right)}\\
u_{n,j,2} & = & \frac{2\omega p_{2}\left(f_{j}^{n-N-1}+f_{j}^{-n}\right)}{\pi\gamma_{2}\left(f_{j}-f_{j}^{-1}\right)\left(f_{j}^{-N-1}-1\right)}
\end{eqnarray}

To obtain the values of the unknown parameters, we explicitly take
into account the conditions of impacts, that should be enforced when
the particle achieves the barriers:
\begin{equation}
\begin{array}{c}
u_{0}{\left(0\right)}=\cfrac{\omega\left(f_{0}^{-N-1}+1\right)\left(p_{2}-p_{1}\right)}{\pi\gamma_{2}\left(f_{0}-f_{0}^{-1}\right)\left(f_{0}^{-N-1}-1\right)}-\\
-\sum_{j=1}^{\infty}{\cfrac{2\omega\left(f_{j}^{-N-1}+1\right)p_{1}}{\pi\gamma_{2}\left(f_{j}-f_{j}^{-1}\right)\left(f_{j}^{-N-1}-1\right)}\cos{\left(j\omega\phi\right)}}+\\
+\sum_{j=1}^{\infty}{\cfrac{2\omega\left(f_{j}^{-N-1}+1\right)p_{2}}{\pi\gamma_{2}\left(f_{j}-f_{j}^{-1}\right)\left(f_{j}^{-N-1}-1\right)}}=1+a
\end{array}
\end{equation}
\begin{equation}
\begin{array}{c}
u_{0}{\left(\phi\right)}=\cfrac{\omega\left(f_{0}^{-N-1}+1_{0}\right)\left(p_{2}-p_{1}\right)}{\pi\gamma_{2}\left(f_{0}-f_{0}^{-1}\right)\left(f_{0}^{-N-1}-1\right)}-\\
-\sum_{j=1}^{\infty}{\cfrac{2\omega\left(f_{j}^{-N-1}+1\right)p_{1}}{\pi\gamma_{2}\left(f_{j}-f_{j}^{-1}\right)\left(f_{j}^{-N-1}-1\right)}}+\\
+\sum_{j=1}^{\infty}{\cfrac{2\omega\left(f_{j}^{-N-1}+1\right)p_{2}}{\pi\gamma_{2}\left(f_{j}-f_{j}^{-1}\right)\left(f_{j}^{-N-1}-1\right)}\cos{\left(j\omega\phi\right)}}=\\
=-1+a
\end{array}
\end{equation}

Reordering these equations, one can write them down in a somewhat
simplified form:
\begin{eqnarray}
-p_{1}\chi_{1}{\left(\phi\right)}+p_{2}\chi_{2} & = & 1+a\label{eq:22}\\
-p_{1}\chi_{2}+p_{2}\chi_{1}{\left(\phi\right)} & = & -1+a\label{eq:23}
\end{eqnarray}
where
\begin{equation}
\chi_{1}{\left(\phi\right)}\equiv\cfrac{\omega}{\pi\gamma_{2}}\left(\begin{array}{c}
\frac{\left(f_{0}^{-N-1}+1\right)}{\left(f_{0}-f_{0}^{-1}\right)\left(f_{0}^{-N-1}-1\right)}+\\
+\sum_{j=1}^{\infty}{\frac{2\left(f_{j}^{-N-1}+1\right)}{\left(f_{j}-f_{j}^{-1}\right)\left(f_{j}^{-N-1}-1\right)}\cos{\left(j\omega\phi\right)}}
\end{array}\right)
\end{equation}
\begin{equation}
\chi_{2}\equiv\cfrac{\omega}{\pi\gamma_{2}}\left(\begin{array}{c}
\frac{\left(f_{0}^{-N-1}+1\right)}{\left(f_{0}-f_{0}^{-1}\right)\left(f_{0}^{-N-1}-1\right)}+\\
+\sum_{j=1}^{\infty}{\frac{2\left(f_{j}^{-N-1}+1\right)}{\left(f_{j}-f_{j}^{-1}\right)\left(f_{j}^{-N-1}-1\right)}}
\end{array}\right)
\end{equation}

So far there are 3 unknowns ($p_{1}$, $p_{2}$ and $\phi$) and only
2 equations. Additional equation is derived from the condition of
energy conservation in the course of each impact:
\begin{equation}
\dot{u}_{0}{\left(0\right)}=-\omega\sum_{j=1}^{\infty}{u_{0,j,1}j\sin{\left(j\omega\phi\right)}}=0
\end{equation}

Due to orthogonality of $\sin{\left(j\omega\phi\right)}$, these conditions
can hold only for $\phi=\pi/\omega$. In other terms, the DB solution
turns out to be symmetric with respect to the time inversion. Fortunately,
this conclusion crucially simplifies the problem as the expression
for $\chi_{1}$ takes the following form:
\begin{equation}
\chi_{1}=\cfrac{\omega}{\pi\gamma_{2}}\left(\begin{array}{c}
\frac{\left(f_{0}^{-N-1}+1\right)}{\left(f_{0}-f_{0}^{-1}\right)\left(f_{0}^{-N-1}-1\right)}+\\
+\sum_{j=1}^{\infty}{\frac{2\left(-1\right)^{j}\left(f_{j}^{-N-1}+1\right)}{\left(f_{j}-f_{j}^{-1}\right)\left(f_{j}^{-N-1}-1\right)}}
\end{array}\right)
\end{equation}

With $\phi$ no longer an unknown, eq. (\ref{eq:22})-(\ref{eq:23})
can now easily be solved:
\begin{eqnarray}
p_{1} & = & \cfrac{1}{\chi_{2}-\chi_{1}}-\cfrac{a}{\chi_{1}+\chi_{2}}\\
p_{2} & = & \cfrac{1}{\chi_{2}-\chi_{1}}+\cfrac{a}{\chi_{1}+\chi_{2}}
\end{eqnarray}

Summarizing, we obtain the following exact solution for the conservative
DB: {\footnotesize{}{}
\begin{equation}
\begin{array}{c}
u_{n}{\left(t\right)}=\cfrac{4\omega\left(f_{0}^{n-N-1}+f_{0}^{-n}\right)a}{\pi\gamma_{2}\left(f_{0}-f_{0}^{-1}\right)\left(f_{0}^{-N-1}-1\right)\left(\chi_{1}+\chi_{2}\right)}+\\
+\underset{j=1}{\overset{\infty}{\sum}}{\left(\cfrac{8\omega a\left(f_{2j}^{n-N-1}+f_{2j}^{-n}\right)}{\pi\gamma_{2}\left(f_{2j}-f_{2j}^{-1}\right)\left(f_{2j}^{-N-1}-1\right)\left(\chi_{1}+\chi_{2}\right)}\right)\cos{\left(j\omega t\right)}}+\\
+\underset{j=1}{\overset{\infty}{\sum}}{\left(\cfrac{8\omega\left(f_{2j-1}^{n-N-1}+f_{2j-1}^{-n}\right)}{\pi\gamma_{2}\left(f_{2j-1}-f_{2j-1}^{-1}\right)\left(f_{2j-1}^{-N-1}-1\right)\left(\chi_{2}-\chi_{1}\right)}\right)\cos{\left(j\omega t\right)}}
\end{array}
\end{equation}
}{\footnotesize \par}

\subsubsection{Single-Sided Discrete Breathers}

The asymmetric barriers allow a new type of DB solution \textendash{}
the single-sided DB, i.e. the regime, in which the impacting mass
does not reach more distant barrier and impacts only one of the barriers.
Consequently, this regime can be described by the following equations
of motion:
\begin{equation}
\ddot{u}_{0}+\gamma_{1}u_{0}+\gamma_{2}\left(2u_{0}-u_{1}-u_{N}\right)=2p\sum_{j=-\infty}^{\infty}{\delta{\left(t-\cfrac{2\pi j}{\omega}\right)}}
\end{equation}
\begin{equation}
\ddot{u}_{n}+\gamma_{1}u_{n}+\gamma_{2}\left(2u_{n}-u_{n+1}-u_{n-1}\right)=0
\end{equation}
\begin{equation}
\ddot{u}_{N}+\gamma_{1}u_{N}+\gamma_{2}\left(2u_{N}-u_{0}-u_{N-1}\right)=0
\end{equation}

As previously, the system is closed with the help of equations, that
fix the impact at the desired location:
\begin{equation}
u_{0}{\left(0\right)}=-1+a
\end{equation}

Finally, the solution is written as follows:

\begin{equation}
\begin{array}{c}
u_{n}=-\cfrac{\omega\left(1-a\right)\left(f_{0}^{n-N-1}+f_{0}^{-n}\right)}{\pi\gamma_{2}\chi_{2}\left(f_{0}-f_{0}^{-1}\right)\left(f_{0}^{-N-1}-1\right)}-\\
-\sum_{j=1}^{\infty}{\cfrac{2\omega\left(1-a\right)\left(f_{j}^{n-N-1}+f_{j}^{-n}\right)}{\pi\gamma_{2}\chi_{2}\left(f_{j}-f_{j}^{-1}\right)\left(f_{j}^{-N-1}-1\right)}\cos{\left(j\omega t\right)}}
\end{array}
\end{equation}
The complete derivation of the solution is available in Appendix \ref{sec:Single-Sided-DB-Derivation}.

\subsection{Discrete Breathers in the Forced-Damped Setting\label{sec:Forced-Damped-Model}}

Now let us adopt that all masses are subjected to an external force
$F{\left(t\right)}$. We examine the case of a symmetric force $F{\left(t\right)}$
which satisfies $F{\left(t\right)}=F{\left(t+2\pi/\omega\right)}$
and $F{\left(t\right)}=-F{\left(t+\pi/\omega\right)}$. Additionally,
the damping is introduced through the non-unit restitution coefficient
$0<e<1$. Similarly to the conservative case, we look for the periodic
solution, thus the impacts can be introduced in the same form as above.
The solution should obey the following set of equations:
\begin{equation}
\begin{array}{c}
\ddot{v}_{0}+\gamma_{1}v_{0}+\gamma_{2}\left(2v_{0}-v_{1}-v_{N}\right)=F{\left(t+\psi\right)}+\\
+2p_{1}\sum_{j=-\infty}^{\infty}{\delta{\left(t-\phi-\cfrac{2\pi j}{\omega}\right)}}-\\
-2p_{2}\sum_{j=-\infty}^{\infty}{\delta{\left(t-\cfrac{2\pi j}{\omega}\right)}}
\end{array}
\end{equation}
\begin{equation}
\ddot{v}_{n}+\gamma_{1}v_{n}+\gamma_{2}\left(2v_{n}-v_{n+1}-v_{n-1}\right)=F{\left(t+\psi\right)}
\end{equation}
\begin{equation}
\ddot{v}_{N}+\gamma_{1}v_{N}+\gamma_{2}\left(2v_{N}-v_{0}-v_{N-1}\right)=F{\left(t+\psi\right)}
\end{equation}
where $\psi$ is the phase of the external force with respect to DB's
impacts.

The external force $F{\left(t\right)}$ can be removed from the equations
with the help of a simple transformation. Let $v_{n}{\left(t\right)}=u_{n}{\left(t\right)}+G{\left(t+\psi\right)}$
where $\ddot{G}{\left(t\right)}+\gamma_{1}G{\left(t\right)}=F{\left(t\right)}$.
Substitution into the above equations yields:
\begin{equation}
\begin{array}{c}
\ddot{u}_{0}+\gamma_{1}u_{0}+\gamma_{2}\left(2u_{0}-u_{1}-u_{N}\right)=\\
=2p_{1}\sum_{j=-\infty}^{\infty}{\delta{\left(t-\phi-\cfrac{2\pi j}{\omega}\right)}}-\\
-2p_{2}\sum_{j=-\infty}^{\infty}{\delta{\left(t-\cfrac{2\pi j}{\omega}\right)}}
\end{array}
\end{equation}
\begin{equation}
\ddot{u}_{n}+\gamma_{1}u_{n}+\gamma_{2}\left(2u_{n}-u_{n+1}-u_{n-1}\right)=0
\end{equation}
\begin{equation}
\ddot{u}_{N}+\gamma_{1}u_{N}+\gamma_{2}\left(2u_{N}-u_{0}-u_{N-1}\right)=0
\end{equation}

Similarly, the impact forcing terms are re-written in the form of
generalized Fourier series:

\begin{equation}
\begin{array}{c}
\ddot{u}_{0}+\gamma_{1}u_{0}+\gamma_{2}\left(2u_{0}-u_{1}-u_{N}\right)=\\
=\frac{\omega}{\pi}\sum_{j=-\infty}^{\infty}{\left(p_{1}\cos{\left(j\omega\left(t-\phi\right)\right)}-p_{2}\cos{\left(j\omega t\right)}\right)}
\end{array}
\end{equation}
\begin{equation}
\ddot{u}_{n}+\gamma_{1}u_{n}+\gamma_{2}\left(2u_{n}-u_{n+1}-u_{n-1}\right)=0
\end{equation}
\begin{equation}
\ddot{u}_{N}+\gamma_{1}u_{N}+\gamma_{2}\left(2u_{N}-u_{0}-u_{N-1}\right)=0
\end{equation}

The equations are identical to those of the conservative model. Hence,
the solution is similar:

\begin{equation}
u_{n}=u_{n,0}+\sum_{j=1}^{\infty}{\left(u_{n,j,1}\cos{\left(j\omega\left(t-\phi\right)\right)}+u_{n,j,2}\cos{\left(j\omega t\right)}\right)}
\end{equation}

where
\begin{eqnarray}
u_{n,0} & = & \frac{\omega\left(p_{2}-p_{1}\right)\left(f_{0}^{n-N-1}+f_{0}^{-n}\right)}{\pi\gamma_{2}\left(f_{0}-f_{0}^{-1}\right)\left(f_{0}^{-N-1}-1\right)}\\
u_{n,j,1} & = & -\frac{2\omega p_{1}\left(f_{j}^{n-N-1}+f_{j}^{-n}\right)}{\pi\gamma_{2}\left(f_{j}-f_{j}^{-1}\right)\left(f_{j}^{-N-1}-1\right)}\\
u_{n,j,2} & = & \frac{2\omega p_{2}\left(f_{j}^{n-N-1}+f_{j}^{-n}\right)}{\pi\gamma_{2}\left(f_{j}-f_{j}^{-1}\right)\left(f_{j}^{-N-1}-1\right)}
\end{eqnarray}

As in the conservative setting, the solution must satisfy the impact
location equations:
\begin{equation}
v_{0}{\left(0\right)}=-p_{1}\chi_{1}{\left(\phi\right)}+p_{2}\chi_{2}+G{\left(\psi\right)}=1+a\label{eq:47}
\end{equation}
\begin{equation}
v_{0}{\left(\phi\right)}=-p_{1}\chi_{2}+p_{2}\chi_{1}{\left(\phi\right)}+G{\left(\psi+\phi\right)}=-1+a\label{eq:48}
\end{equation}

Also, the impact law must be satisfied:
\begin{equation}
\begin{array}{c}
\dot{v}_{0}{\left(0^{+}\right)}=\dot{u}_{0}{\left(0^{+}\right)}+\dot{G}{\left(\psi\right)}=\\
=-e\left(\dot{u}_{0}{\left(0^{-}\right)}+\dot{G}{\left(\psi\right)}\right)=-e\dot{v}_{0}{\left(0^{-}\right)}
\end{array}
\end{equation}

\begin{equation}
\dot{u}_{0}{\left(0^{+}\right)}+e\dot{u}_{0}{\left(0^{-}\right)}=-\dot{G}{\left(\psi\right)}\left(1+e\right)\label{eq:71}
\end{equation}

For the symmetric case it is clear that the energy must be conserved
during each impact in terms of $u_{0}$, namely, in terms of the reduced
un-forced system. Hence, for the un-forced system and the symmetric
DB $\dot{u}_{0}{\left(0^{+}\right)}=-\dot{u}_{0}{\left(0^{-}\right)}$
and similarly for the second impact. However, this is not true for
the asymmetric DB \textendash{} the energy must be conserved for the
reduced un-forced system (otherwise, the DB solution cannot exist),
but it holds for the complete period of oscillations, and not necessarily
in each single impact. So, more refined treatment is required in this
case.

The generalized Fourier series converges to the average of the velocities
on both sides of the discontinuity:

\begin{equation}
\cfrac{\dot{u}_{0}{\left(0^{+}\right)}+\dot{u}_{0}{\left(0^{-}\right)}}{2}=-p_{1}\chi_{3}
\end{equation}
where
\begin{equation}
\chi_{3}\equiv\sum_{j=1}^{\infty}{\frac{2j\omega^{2}\left(f_{j}^{-N-1}+1\right)}{\pi\gamma_{2}\left(f_{j}-f_{j}^{-1}\right)\left(f_{j}^{-N-1}-1\right)}\sin{\left(j\omega\phi\right)}}
\end{equation}

Conservation of momentum during the impact yields:
\begin{equation}
\cfrac{\dot{u}_{0}{\left(0^{+}\right)}-\dot{u}_{0}{\left(0^{-}\right)}}{2}=-p_{2}
\end{equation}

From these equations we extract terms for the velocities:
\begin{equation}
\dot{u}_{0}{\left(0^{+}\right)}=-p_{2}-p_{1}\chi_{3}
\end{equation}
\begin{equation}
\dot{u}_{0}{\left(0^{-}\right)}=p_{2}-p_{1}\chi_{3}
\end{equation}

Note that the energy gain for the reduced un-forced system during
the impact is $\Delta E_{1}=2p_{1}p_{2}\chi_{3}$. Energy gain is
possible since the reduced un-forced system does not represent a physical
system.

Plugging into eq. (\ref{eq:71}), one obtains:
\begin{equation}
\dot{G}{\left(\psi\right)}=p_{1}\chi_{3}+qp_{2}\label{eq:75}
\end{equation}
where $q=\left(1-e\right)/\left(1+e\right)$.

Similarly, it is possible to perform the same procedure for the second
impact:

\begin{equation}
\dot{u}_{0}{\left(\phi^{+}\right)}+e\dot{u}{\left(\phi^{-}\right)}=-\dot{G}{\left(\phi+\psi\right)}\left(1+e\right)\label{eq:78}
\end{equation}

The generalized Fourier series converges to the average of the velocities
on both sides of the discontinuity:

\begin{equation}
\cfrac{\dot{u}{\left(\phi^{+}\right)}+\dot{u}{\left(\phi^{-}\right)}}{2}=-p_{2}\chi_{3}
\end{equation}

Conservation of momentum during the impact yields:
\begin{equation}
\cfrac{\dot{u}{\left(\phi^{+}\right)}-\dot{u}{\left(\phi^{-}\right)}}{2}=p_{1}
\end{equation}

From these equations we extract the terms for the velocities:
\begin{equation}
\dot{u}{\left(\phi^{+}\right)}=p_{1}-p_{2}\chi_{3}
\end{equation}
\begin{equation}
\dot{u}{\left(\phi^{-}\right)}=-p_{1}-p_{2}\chi_{3}
\end{equation}

Note that the energy loss in this impact is $\Delta E_{2}=-2p_{1}p_{2}\chi_{3}$;
hence the energy of the reduced un-forced system is conserved throughout
the period as expected.

Plugging into eq. (\ref{eq:78}), one obtains:
\begin{equation}
\dot{G}{\left(\phi+\psi\right)}=p_{2}\chi_{3}-qp_{1}\label{eq:81}
\end{equation}

\subsubsection{Harmonic Excitation}

In order to solve the equations we need to choose the forcing function,
that satisfies the symmetry conditions. For simplicity, let us choose
$F{\left(t\right)}=A\cos{\left(\omega t\right)}$. Solving the ODE,
we obtain:
\begin{equation}
G{\left(t\right)}=\tilde{A}\cos{\left(\omega t\right)}
\end{equation}
where,
\begin{equation}
\tilde{A}=\cfrac{A}{\gamma_{1}-\omega^{2}}
\end{equation}

Plugging the solution into eq. (\ref{eq:47}), (\ref{eq:48}), (\ref{eq:75})
and (\ref{eq:81}), one obtains the following expressions:
\begin{equation}
-p_{1}\chi_{1}{\left(\phi\right)}+p_{2}\chi_{2}+\tilde{A}\cos{\left(\omega\psi\right)}=1+a
\end{equation}
\begin{equation}
-p_{1}\chi_{2}+p_{2}\chi_{1}{\left(\phi\right)}+\tilde{A}\cos{\left(\omega\left(\psi+\phi\right)\right)}=-1+a
\end{equation}

\begin{equation}
-\tilde{A}\omega\sin{\left(\omega\left(\psi+\phi\right)\right)}=p_{2}\chi_{3}-qp_{1}
\end{equation}
\begin{equation}
-\tilde{A}\omega\sin{\left(\omega\psi\right)}=p_{1}\chi_{3}+qp_{2}
\end{equation}

To find the exact solution explicitly, we assume that $\phi$ is known
and the barrier asymmetry $a$ is the unknown. Solution of the above
set of equations under this assumption yields:
\begin{equation}
\psi=\cfrac{\pm\arccos{\left(\cfrac{2\left(q^{2}+\chi_{3}^{2}\right)}{\sigma\tilde{A}}\right)}+\alpha}{\omega}
\end{equation}
\begin{equation}
p_{1}=\cfrac{\tilde{A}\omega\left(q\sin{\left(\omega\left(\psi+\phi\right)\right)}-\chi_{3}\sin{\left(\omega\psi\right)}\right)}{q^{2}+\chi_{3}^{2}}
\end{equation}
\begin{equation}
p2=-\cfrac{\tilde{A}\omega\left(\chi_{3}\sin{\left(\omega\left(\psi+\phi\right)\right)}+q\sin{\left(\omega\psi\right)}\right)}{q^{2}+\chi_{3}^{2}}
\end{equation}
\begin{equation}
a=-p_{1}\chi_{1}{\left(\phi\right)}+p_{2}\chi_{2}+\tilde{A}\cos{\left(\omega\psi\right)}-1
\end{equation}

\begin{widetext}
where,
\begin{equation}
\sigma=\sqrt{\begin{array}{c}
2\omega^{2}\chi_{3}^{2}\left(\chi_{1}-\chi_{2}\right)^{2}\left(1+\cos{\left(\omega\phi\right)}\right)+4\omega\chi_{3}\left(q^{2}+\chi_{3}^{2}\right)\left(\chi_{1}-\chi_{2}\right)\sin{\left(\omega\phi\right)}+\\
+2\left(q^{4}+\chi_{3}^{4}+q^{2}\left(\omega^{2}\left(\chi_{1}-\chi_{2}\right)^{2}+2\chi^{3}\right)\right)\left(1-\cos{\left(\omega\phi\right)}\right)
\end{array}}
\end{equation}
\begin{equation}
\alpha=\pm\arccos{\left(\cfrac{\left(q^{2}+\chi_{3}^{2}\right)\left(1-\cos{\left(\omega\phi\right)}\right)-\left(q-\chi_{3}\right)\left(\chi_{1}-\chi_{2}\right)\omega\sin{\left(\omega\phi\right)}}{\sigma}\right)}
\end{equation}
\end{widetext}

\subsubsection{Single-Sided Forced-Damped Discrete Breathers}

The single-sided DB is also possible in the forced-damped model. The
equations of motion can be written as follows:
\begin{equation}
\begin{array}{c}
\ddot{v}_{0}+\gamma_{1}v_{0}+\gamma_{2}\left(2v_{0}-v_{1}-v_{N}\right)=\\
=F{\left(t+\psi\right)}+2p\sum_{j=-\infty}^{\infty}{\delta{\left(t-\cfrac{2\pi j}{\omega}\right)}}
\end{array}
\end{equation}
\begin{equation}
\ddot{v}_{n}+\gamma_{1}v_{n}+\gamma_{2}\left(2v_{n}-v_{n+1}-v_{n-1}\right)=F{\left(t+\psi\right)}
\end{equation}
\begin{equation}
\ddot{v}_{N}+\gamma_{1}v_{N}+\gamma_{2}\left(2v_{N}-v_{0}-v_{N-1}\right)=F{\left(t+\psi\right)}
\end{equation}
where $\psi$ is the phase of the external force with respect to the
DB's impacts.

The external force $F{\left(t\right)}$ can be removed from the equations
in the same manner as the previous case.

As in the conservative model, the solution must satisfy the impact
location equations:
\begin{equation}
v_{0}{\left(0\right)}=-p\chi_{2}+G{\left(\psi\right)}=-1+a\label{eq:48-2}
\end{equation}

Also, the impact law must be satisfied:
\begin{equation}
\begin{array}{c}
\dot{v}{\left(0^{+}\right)}=\dot{u}{\left(0^{+}\right)}+\dot{G}{\left(\psi\right)}=p+\dot{G}{\left(\psi\right)}=\\
=-e\left(-p+\dot{G}{\left(\psi\right)}\right)=-e\left(\dot{u}{\left(0^{-}\right)}+\dot{G}{\left(\psi\right)}\right)=-e\dot{v}{\left(0^{-}\right)}
\end{array}
\end{equation}

By further simplification, one obtains:
\begin{eqnarray}
\dot{G}{\left(\psi\right)} & = & -qp\label{eq:51-1}
\end{eqnarray}

The full derivations are given in Appendix\ref{sec:Single-Sided-Forced-DB-Derivations}.

\paragraph{Harmonic Excitation}

Let $F{\left(t\right)}=A\cos{\left(\omega t\right)}$ and,
\begin{equation}
G{\left(t\right)}=\tilde{A}\cos{\left(\omega t\right)}
\end{equation}
where,
\begin{equation}
\tilde{A}=\cfrac{A}{\gamma_{1}-\omega^{2}}
\end{equation}

In a similar manner to the regular DB, we obtain the following equations:
\begin{equation}
-p\chi_{2}+\tilde{A}\cos{\left(\omega\psi\right)}=-1+a\label{eq:48-2-1}
\end{equation}

\begin{eqnarray}
-\omega\tilde{A}\sin{\left(\omega\psi\right)} & = & -qp\label{eq:51-1-1}
\end{eqnarray}
Solving the equation yields:
\begin{equation}
\psi=\cfrac{\pm\arccos{\left(\cfrac{q\left(a-1\right)}{\tilde{A}\sqrt{q^{2}+\omega^{2}\chi_{2}^{2}}}\right)}+\alpha}{\omega}
\end{equation}
\begin{equation}
p=\cfrac{\omega\tilde{A}}{q}\sin{\left(\omega\psi\right)}
\end{equation}
where,
\begin{equation}
\alpha=\pm\arccos{\left(\cfrac{q}{\sqrt{q^{2}+\omega^{2}\chi_{2}^{2}}}\right)}
\end{equation}

\subsubsection{Single-Sided Forced-Damped Discrete Breathers with Period Doubling}

The stability analysis discussed in detail in the following Section
shows that one of the mechanisms for the loss of stability of the
single-sided forced DB is the period doubling bifurcation. Numerical
investigation shows that the period doubling is reflected by a consecutive
set of collisions with different exchange of momentum in the new doubled
period of the DB. This type of solution can also be obtained analytically
in a similar manner to that of the forced DB with minor modifications.
The location of the second impact in the period is set to the same
barrier as the first collision, i.e. the closer barrier, and the period
of the DB is doubled. However, obtaining a solution in this manner
is only possible if the frequency of the doubled period solution is
in the attenuation zone of the chain for a given set of parameters.
The full derivation is given in detail in Appendix \ref{sec:Period-Doubling-Derivations}.

\section{Stability\label{sec:Stability}}

The stability of the derived DB solutions will be investigated by
Floquet theory\cite{Strogatz1994}. The Floquet multipliers are often
evaluated numerically, but, as mentioned above, the special nature
of the system allows explicit construction of the monodromy matrix.
Then, computation of its eigenvalues is a relatively simple computational
task, and comprehensive study of the stability patterns in the space
of parameters becomes possible \cite{Gendelman2013}. Moreover, eigenvectors
corresponding to the unstable Floquet multipliers can be easily computed
and examined to gain some qualitative insight into the mechanism of
the loss of stability.
\begin{widetext}
The governing equations of motion can also be written in the following
equivalent form:
\begin{equation}
\dot{\vec{u}}=\mbox{A}\vec{u}
\end{equation}
where $\vec{u}=\left[\begin{array}{cccccc}
u_{0} & \cdots & u_{N} & \dot{u}_{0} & \cdots & \dot{u}_{N}\end{array}\right]^{T}$ and:
\begin{eqnarray}
\mbox{A} & = & \left[\begin{array}{cc}
\mbox{0}_{\left(N+1\right)\times\left(N+1\right)} & \mbox{I}_{\left(N+1\right)\times\left(N+1\right)}\\
\tilde{\mbox{A}}_{\left(N+1\right)\times\left(N+1\right)} & \mbox{0}_{\left(N+1\right)\times\left(N+1\right)}
\end{array}\right]\\
\mbox{\ensuremath{\tilde{A}}} & = & \left[\begin{array}{cccccc}
\gamma_{1}+2\gamma_{2} & -\gamma_{2} & 0 & \cdots & 0 & -\gamma_{2}\\
-\gamma_{2} & \gamma_{1}+2\gamma_{2} & -\gamma_{2} & 0 & \cdots & 0\\
0 & -\gamma_{2} & \ddots & \ddots & \ddots & \vdots\\
\vdots & \ddots & \ddots & \gamma_{1}+2\gamma_{2} & -\gamma_{2} & 0\\
0 & \cdots & 0 & -\gamma_{2} & \gamma_{1}+2\gamma_{2} & -\gamma_{2}\\
-\gamma_{2} & 0 & \cdots & 0 & -\gamma_{2} & \gamma_{1}+2\gamma_{2}
\end{array}\right]
\end{eqnarray}
or for the forced-damped model:
\begin{equation}
\dot{\vec{v}}=\mbox{A}\vec{v}+\vec{F}
\end{equation}
where $\vec{F}=F{\left(t\right)}\left[\begin{array}{ccc}
1 & \cdots & 1\end{array}\right]^{T}$.
\end{widetext}

From the above equation we can derive the evolution of the perturbed
phase trajectory between two impacts:
\begin{equation}
\mbox{L}_{1}=\exp{\left(\phi\mbox{A}\right)}
\end{equation}

\begin{equation}
\mbox{L}_{2}=\exp{\left(\left(\cfrac{2\pi}{\omega}-\phi\right)\mbox{A}\right)}
\end{equation}
or for the single-sided impact:
\begin{equation}
\mbox{L}=\exp{\left(\cfrac{2\pi}{\omega}\mbox{A}\right)}
\end{equation}

The impacts mapping cannot simply be based on the impact law for the
stability analysis, but saltation matrix must be constructed to take
into account the linear perturbations of the mapping and of the flight
time to the discontinuity\cite{Fredriksson2000}. The saltation matrix
for the adopted impact law obtains the following form:
\begin{equation}
\mbox{S}_{1,2}=\left[\begin{array}{cc}
\mbox{\ensuremath{\tilde{S}}}_{\left(N+1\right)\times\left(N+1\right)} & \mbox{0}_{\left(N+1\right)\times\left(N+1\right)}\\
\hat{\mbox{S}}_{\left(N+1\right)\times\left(N+1\right)} & \mbox{\ensuremath{\tilde{S}}}_{\left(N+1\right)\times\left(N+1\right)}
\end{array}\right]
\end{equation}
where,
\begin{equation}
\tilde{\mbox{S}}=\left[\begin{array}{ccccc}
-e & 0 & \cdots & \cdots & 0\\
0 & 1 & 0 &  & \vdots\\
\vdots & 0 & \ddots & \ddots & \vdots\\
\vdots & \mbox{} & \ddots & 1 & 0\\
0 & \cdots & \cdots & 0 & 1
\end{array}\right]
\end{equation}
\begin{equation}
\hat{\mbox{S}}_{1,2}=\left[\begin{array}{ccccc}
\cfrac{\left(1+e\right)\Delta_{1,2}}{\Gamma_{1,2}} & 0 & \cdots & \cdots & 0\\
0 & 0 & 0 & \mbox{} & \vdots\\
\vdots & 0 & \ddots & \ddots & \vdots\\
\vdots & \mbox{} & \ddots & 0 & 0\\
0 & \cdots & \cdots & 0 & 0
\end{array}\right]
\end{equation}
where $\Delta_{1}=\ddot{u}_{0}{\left(\phi-\right)}$,$\Delta_{2}=\ddot{u}_{0}{\left(0-\right)}$,
, $\Gamma_{1}=-p_{1}$,$\Gamma_{2}=p_{2}$ for the conservative model;
$\Delta_{1}=\ddot{u}_{0}{\left(\phi-\right)}$,$\Delta_{2}=\ddot{u}_{0}{\left(0-\right)}$,
, $\Gamma_{1}=-p_{1}-p_{2}\chi_{3}+\dot{G}{\left(\phi+\psi\right)}$,$\Gamma_{2}=p_{2}-p_{1}\chi_{3}+\dot{G}{\left(\psi\right)}$
for the forced-damped model. Similarly for the Single-sided DB \textendash{}
$\Delta_{1}=\ddot{u}_{0}{\left(0-\right)}$ and $\Gamma_{1}=-p$ for
the conservative model; $\Delta_{1}=\ddot{u}_{0}{\left(0-\right)}$,
, $\Gamma_{1}=-p+\dot{G}{\left(\psi\right)}$ for the forced-damped
model. Note that for the conservative model the coefficient of restitution
$e$ is set to unity.

The Monodromy matrix can be written compactly as follows:
\begin{equation}
\mbox{M}=\mbox{L}_{1}\mbox{S}_{1}\mbox{L}_{2}\mbox{S}_{2}
\end{equation}

or for the single-sided DB:
\begin{equation}
\mbox{M}=\mbox{L}\mbox{S}_{1}
\end{equation}

Then, the stability of the DB solution is assessed just by easy computation
of this Monodromy matrix and evaluation of its spectrum.

\section{Numerical Validation and Stability Patterns\label{sec:Numerical-Validation}}

\subsection{Conservative Model}

In order to qualitatively examine the properties of the asymmetric
DBs, and to validate the accuracy of our results, we turn to numerical
methods. The simulations in this section are performed using MatLab.
The vibro-impact response was modeled according to the impact law
using event-driven algorithm. The numerical results were in agreement
with the analytical results, as one should expect for the exact solutions.
Thus, one can see these results as illustrations. Unless otherwise
stated, the simulations were done for the following set of parameters:

\begin{equation}
\begin{array}{c}
\begin{array}{ccc}
\gamma_{1}=0.2 & \gamma_{2}=0.1 & \omega=1.5\end{array}\\
\begin{array}{cc}
N=20 & a=0.4\end{array}
\end{array}
\end{equation}

\begin{figure}[H]
\begin{centering}
\includegraphics[width=1\columnwidth]{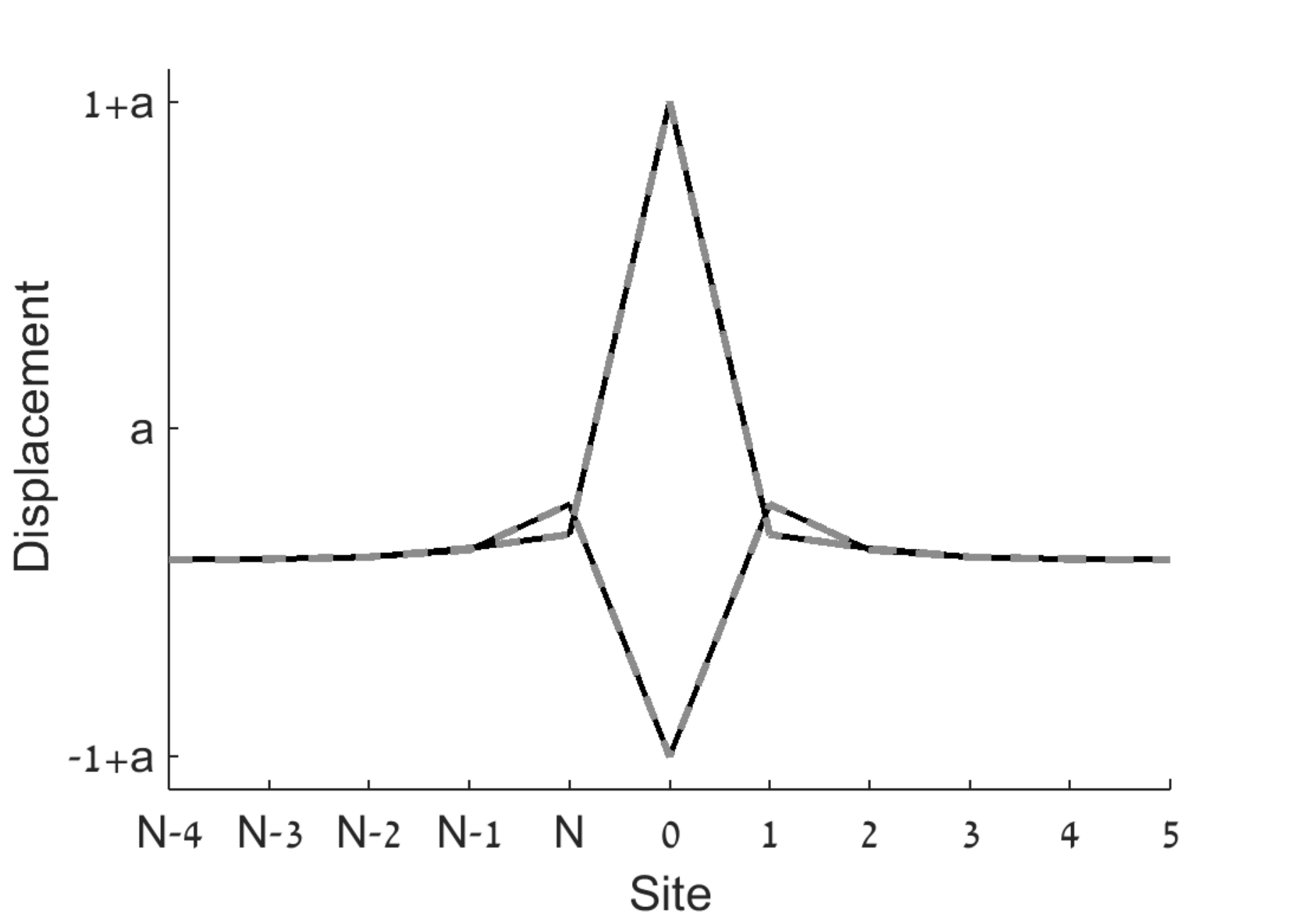}
\par\end{centering}
\caption{\label{fig:1}The displacements of the masses for a DB solution at
the instances of the two impacts for $N=10$ (Black) and $N=100$
(Dashed gray).}
\end{figure}

As was the case in the symmetric DB, the oscillatory profile is qualitatively
the same when the length of the chain is modified, as shown in fig.
\ref{fig:1}. This result conforms to the strong localization of the
DB solution.

\begin{figure}[H]
\begin{centering}
\includegraphics[width=1\columnwidth]{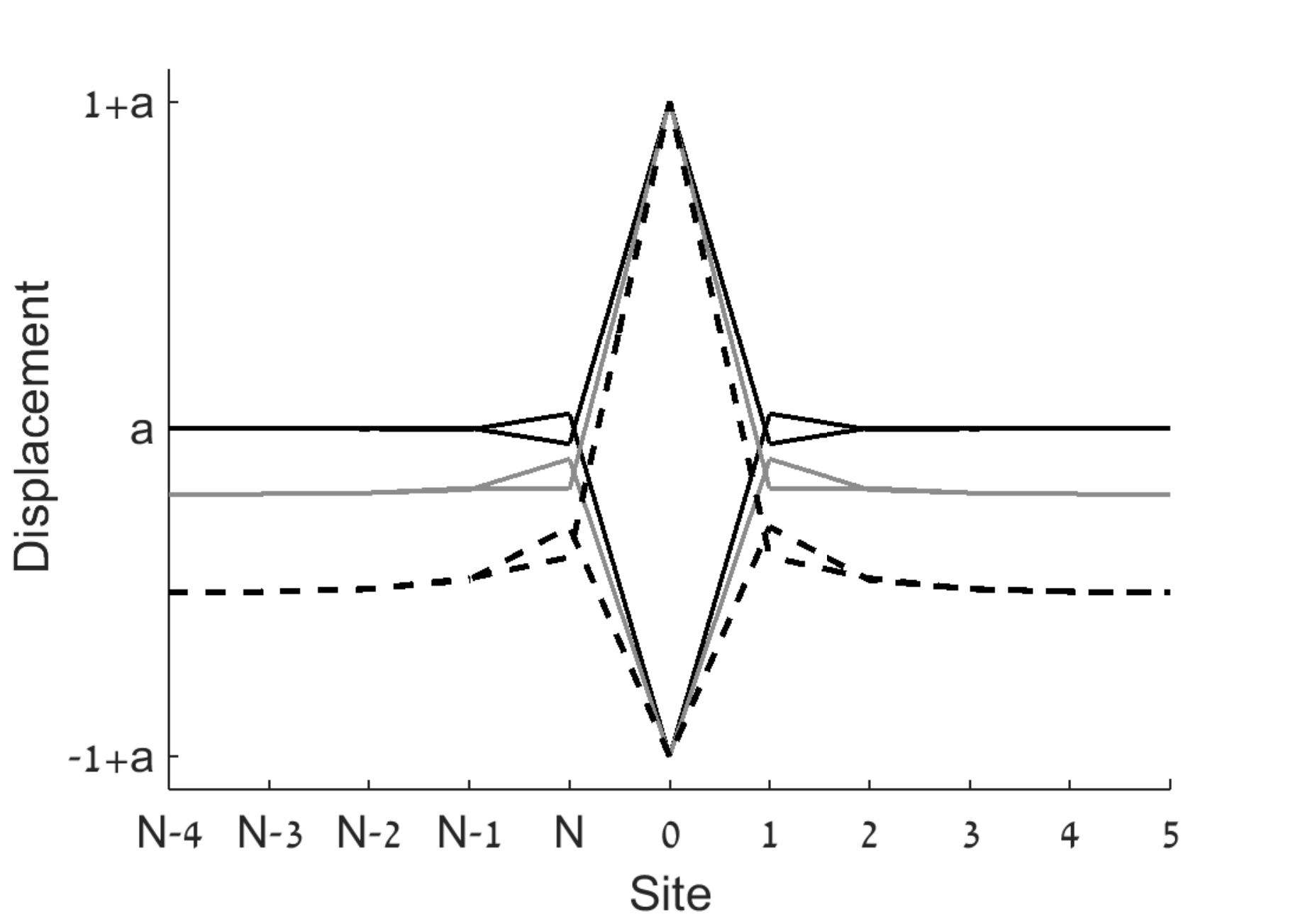}
\par\end{centering}
\caption{\label{fig:2}The displacements of the masses for a DB solution at
the instances of the two impacts for $a=0$ (Black), $a=0.2$ (Gray)
and $a=0.5$ (Dashed Black).}
\end{figure}

\begin{figure}[H]
\begin{centering}
\includegraphics[width=1\columnwidth]{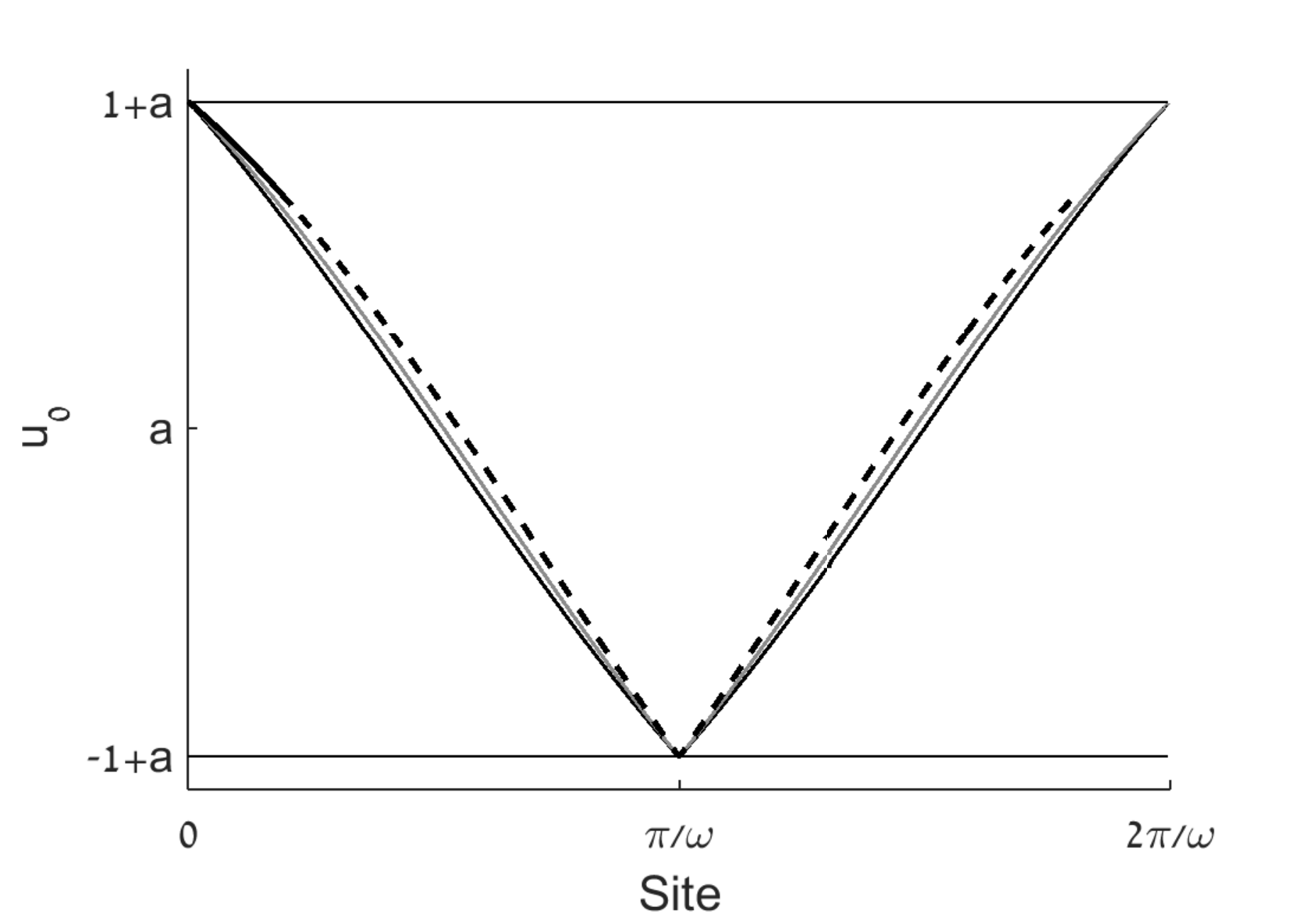}
\par\end{centering}
\caption{\label{fig:3}The displacements of the first mass for a DB solution
for $a=0$ (Black), $a=0.2$ (Gray) and $a=0.5$ (Dashed Black).}
\end{figure}

Figure \ref{fig:2} shows the strong effect of the asymmetry parameter
$a$ on the DB shape. However, when examining the displacement of
the impacting mass throughout the period of the DB , the difference
is only a small change in the curvature as demonstrated in fig. \ref{fig:3}.
It appears that for the conservative DB, the asymmetry mainly contributes
to the value to which the DB converges apart from the localization
site.

\begin{figure}[H]
\begin{centering}
\includegraphics[width=1\columnwidth]{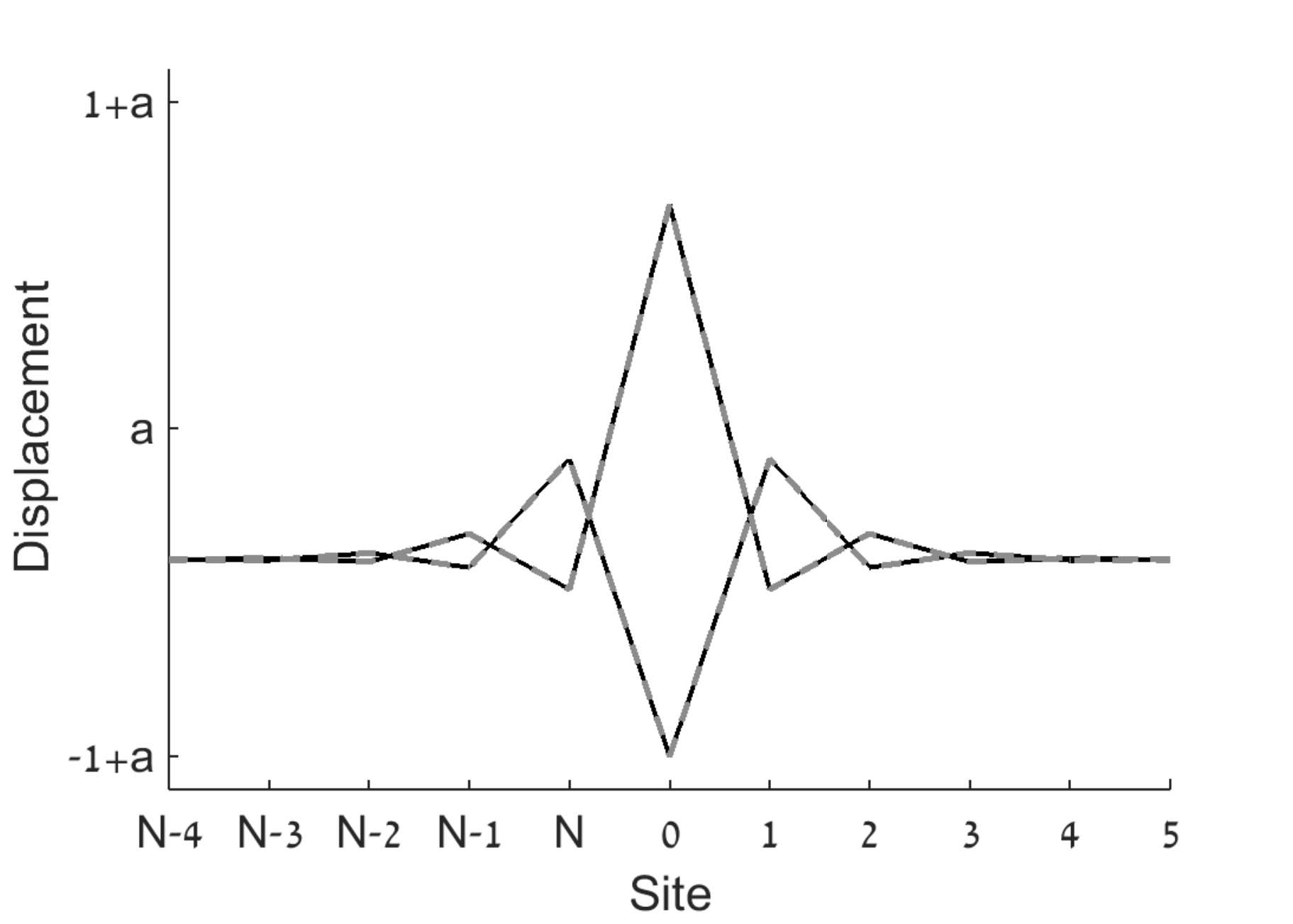}
\par\end{centering}
\caption{\label{fig:4}The displacements of the masses for a single-sided DB
solution at the instances of the two impacts.}
\end{figure}

Another type of DB enabled by the asymmetry of the system is the single-sided
DB. Figure \ref{fig:4} presents the example of the single-sided DB;
note that the impacting mass does not reach the more distant barrier
at $\left(1+a\right)$.

\subsection{Forced-Damped Model}

This model is a bit more complicated to examine. As mentioned in sec.
\ref{sec:Forced-Damped-Model}, we are unable to find $\phi$ without
approximations with unknown error. Therefore, $\phi$ is regarded
as a known parameter and instead we obtain the asymmetry parameter
$a$. Fortunately, numerical investigation reveals that the relation
between $a$ and $\phi$ behaves in a manner that allows finding the
wanted value of $a$ by means of iterative extrapolation with the
maximal error of our choice. For time consumption purposes, the allowed
error in $a$ was taken to be $10^{-12}$. Furthermore, unless stated
otherwise, the parameters are as follows:
\begin{equation}
\begin{array}{c}
\begin{array}{cccc}
\gamma_{1}=0.2 & \gamma_{2}=0.1 & \omega=1.5 & N=20\end{array}\\
\begin{array}{ccc}
A=0.1 & e=0.9 & a=0.4\end{array}
\end{array}
\end{equation}

\begin{figure}[H]
\begin{centering}
\includegraphics[width=1\columnwidth]{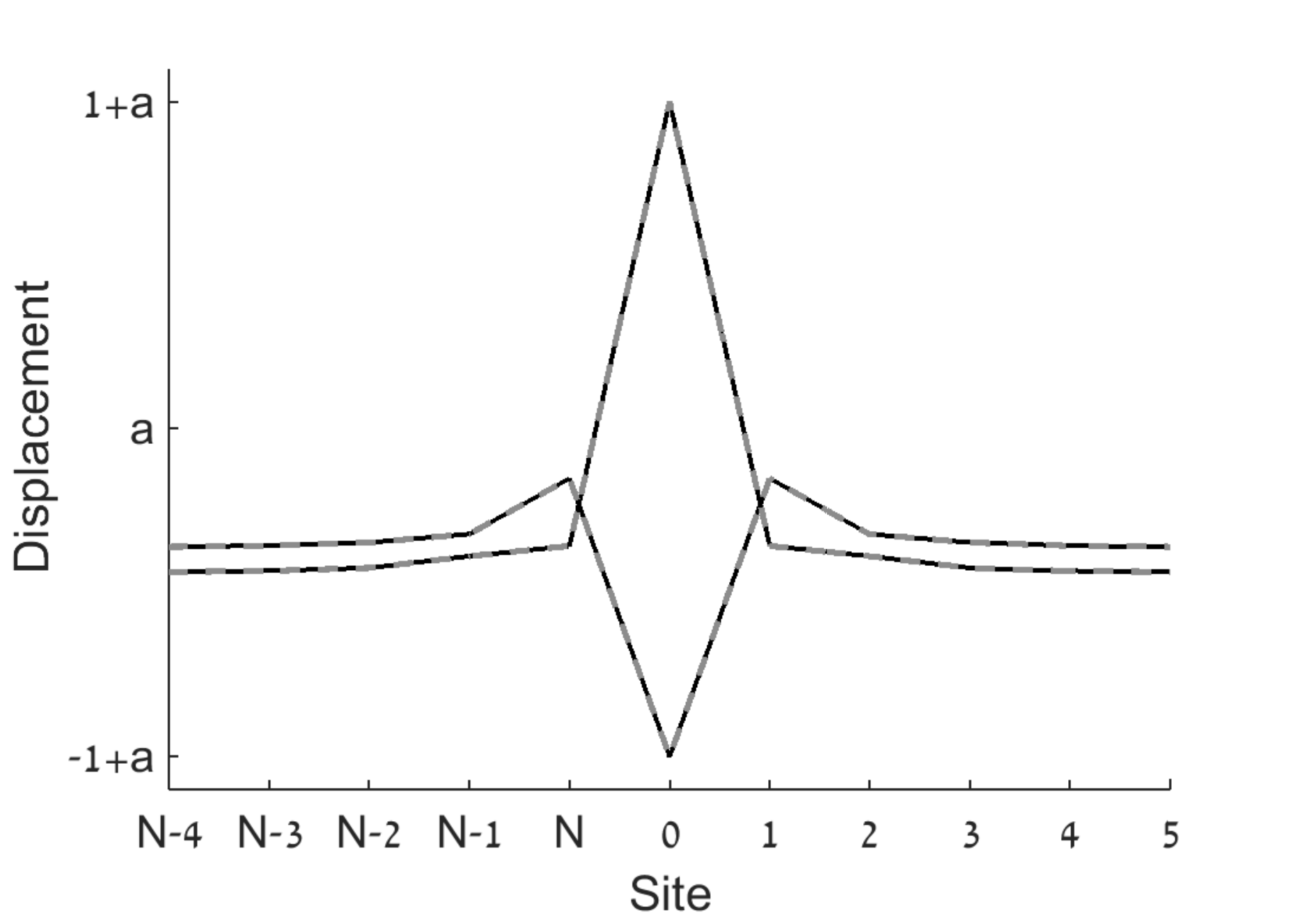}
\par\end{centering}
\caption{\label{fig:5}The displacements of the masses for the forced DB solution
at the instances of the two impacts for $N=10$ (Black) and $N=100$
(Dashed gray).}
\end{figure}

\begin{figure}[H]
\begin{centering}
\includegraphics[width=1\columnwidth]{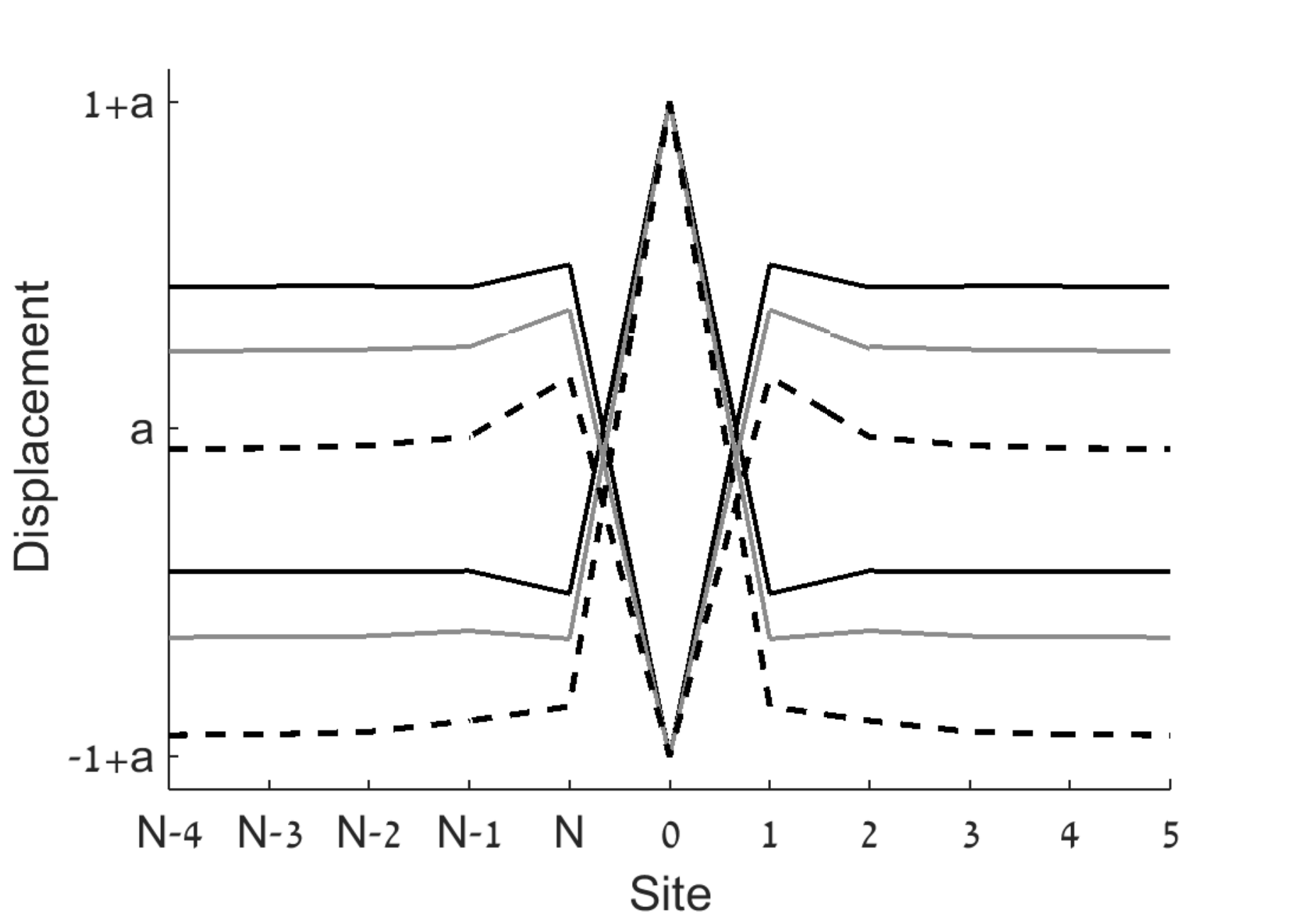}
\par\end{centering}
\caption{\label{fig:6}The displacements of the masses for the forced DB solution
at the instances of the two impacts for $A=0.9$ and $a=0$ (Black),
$a=0.2$ (Gray) and $a=0.4$ (Dashed Black).}
\end{figure}

In general, the effect of the asymmetry in the forced-damped model
is similar to one observed in the conservative model. In fig. \ref{fig:5}
we see that there is no notable change in the DB profile as a result
of adding masses to the chain. Figure \ref{fig:6} shows that, while
the shape is generally different since the oscillating term converges
to $G{\left(t\right)}$ and not to zero, as the mass is farther away
from the localization site, the most profound consequence of the asymmetry
is still the shift of the center of oscillations.

\begin{figure}[H]
\begin{centering}
\includegraphics[width=1\columnwidth]{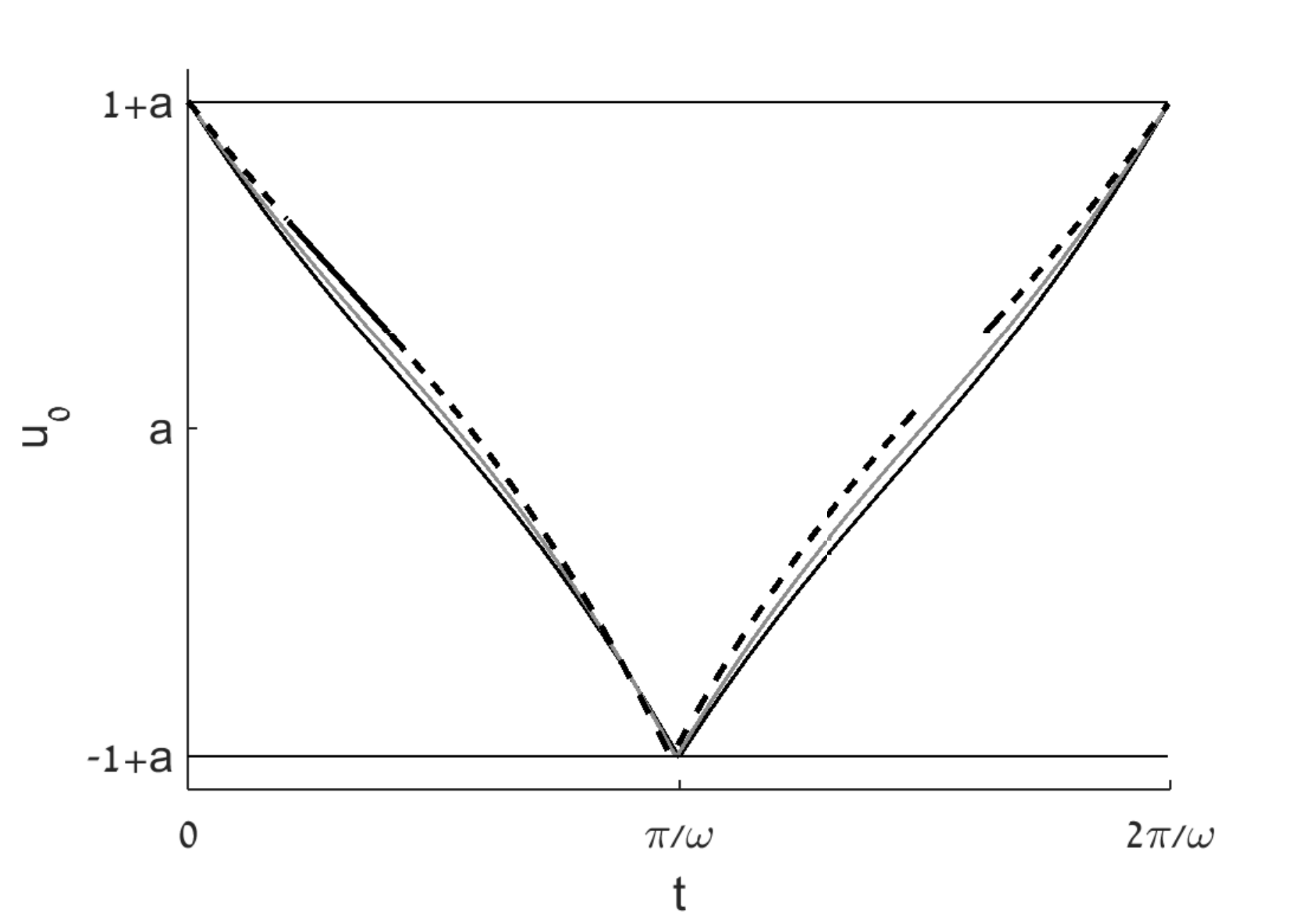}
\par\end{centering}
\caption{\label{fig:7}The displacements of the first mass for $A=0.9$ and
$a=0$ (Black), $a=0.2$ (Gray) and $a=0.5$ (Dashed Black).}
\end{figure}

Figure \ref{fig:7} demonstrates a main difference from the conservative
DB; it clearly shows that there is a shift of the second impact, i.e.
$\phi$ is diverted from $\pi/\omega$. Numerical investigation shows
this difference is typically very small until the appearance of multiple
solutions mentioned below.

\begin{figure}[H]
\begin{centering}
\includegraphics[width=1\columnwidth]{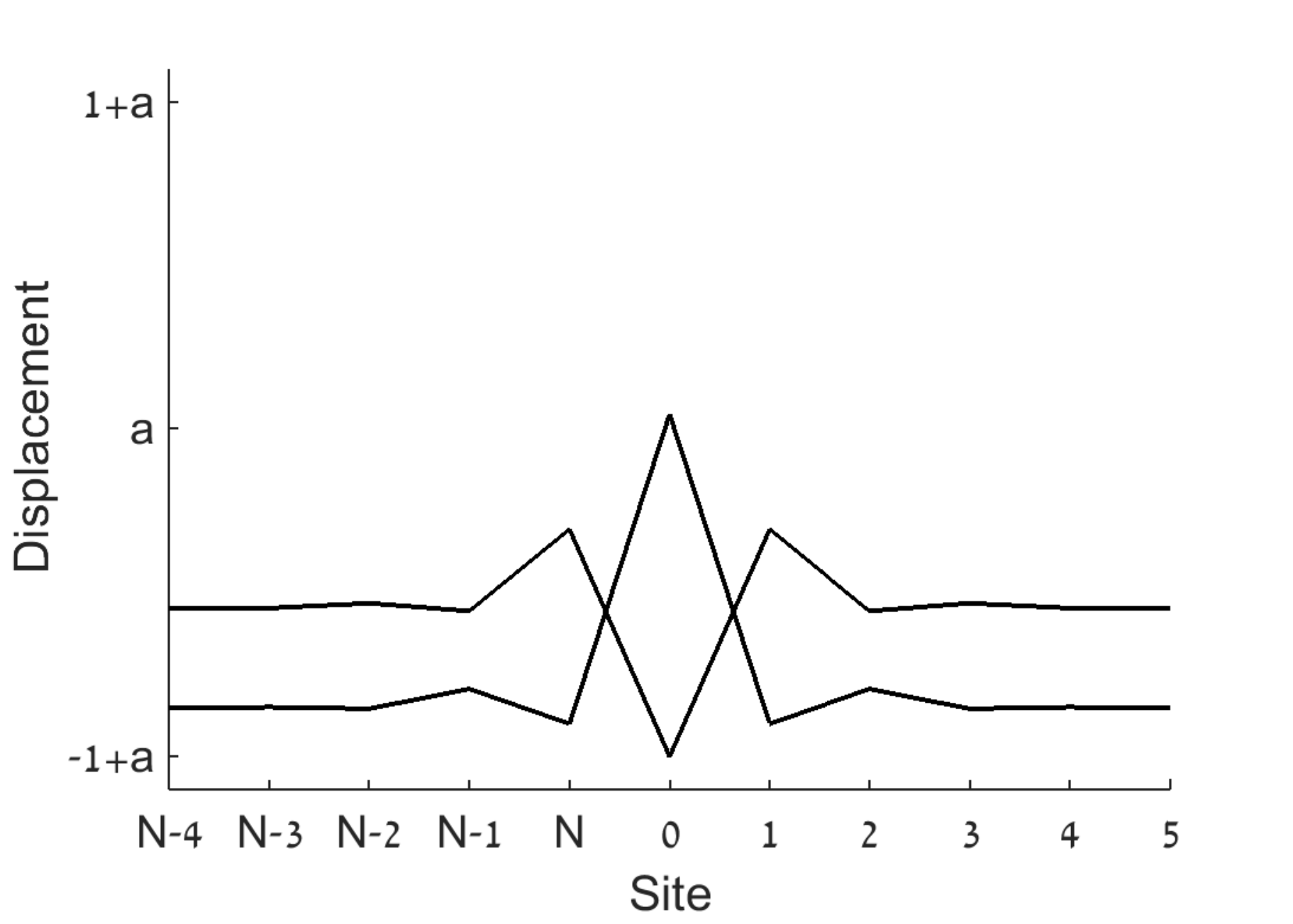}
\par\end{centering}
\caption{\label{fig:8}The displacements of the masses for the forced single-sided
DB solution at the instances of the two impacts for $\omega=0.92$.}
\end{figure}

Just like in the conservative model, the forced single-sided DBs exist
as well. An example is presented in fig. \ref{fig:8}.

\begin{figure}[H]
\begin{centering}
\includegraphics[width=1\columnwidth]{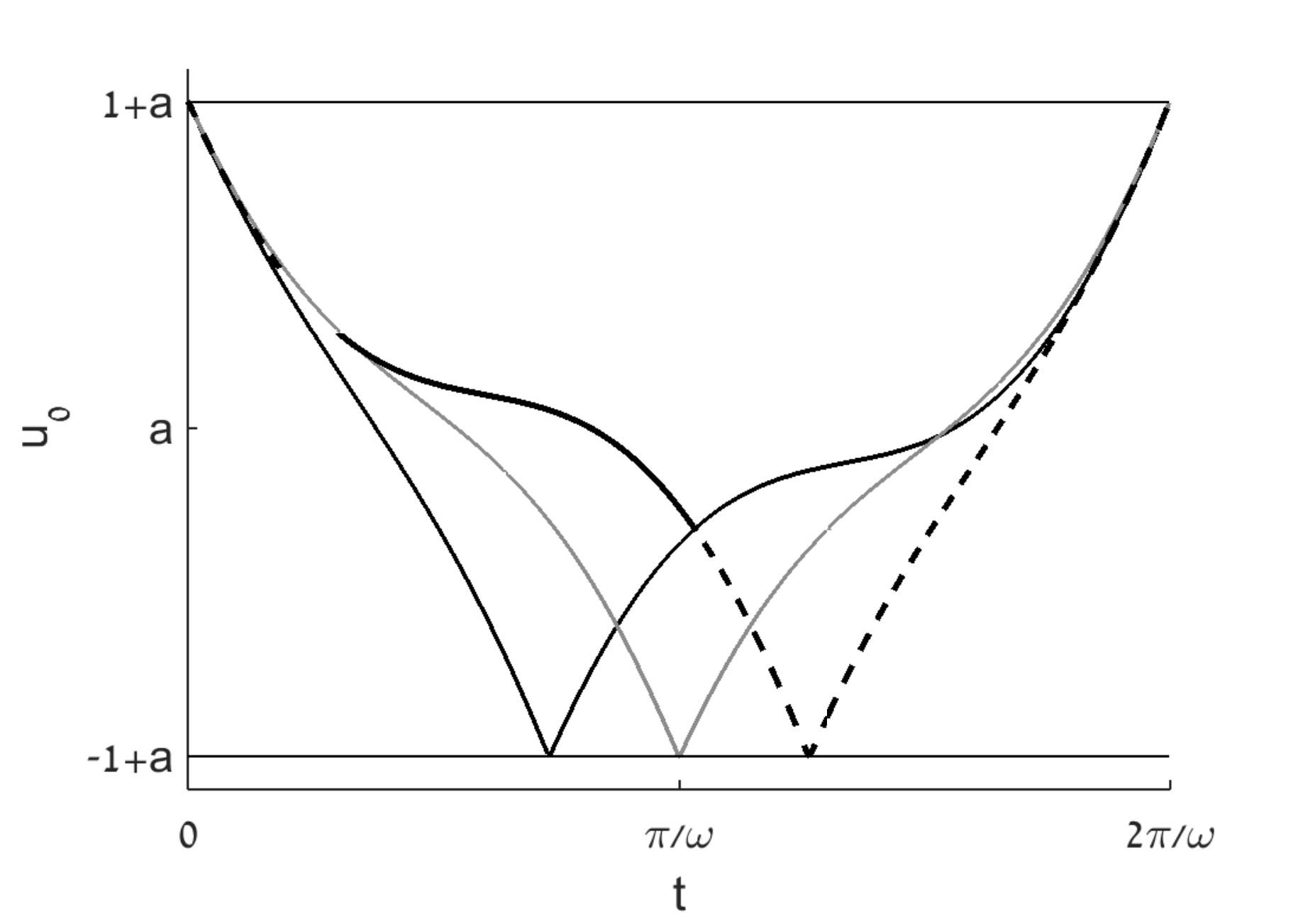}
\par\end{centering}
\caption{\label{fig:9}The displacements of the first mass for some of the
solutions for $\gamma_{1}=0.1$, $\gamma_{2}=0.05$, $\omega=1.33$,
$A=1.5$ and $a=0$.}
\end{figure}

An interesting phenomenon appearing for stronger external forcing,
i.e. larger values of $A$, is a multitude of solutions. For certain
sets of parameters the analytic solution yields more than a single
solution. Figure \ref{fig:9} shows that these solutions can even
be very different from each other. It is interesting to note that
it is possible that more than one of these solutions are stable.

Another interesting fact is that even for a symmetric model, namely
$a=0$, an asymmetric solution could exist as predicted by Grinberg
and Gendelman \cite{Grinberg2016}. one or more of these solutions
appear to become stable when the symmetric DB losses stability via
the pitchfork bifurcation.

\subsection{Stability}

The procedure for the stability analysis is described in detail in
sec. \ref{sec:Stability}. Additionally, the following stability maps
also refer to existence of the solution, i.e. solutions that are not
physical, e.g. some masses exceed the boundaries, are marked as non-existent.
The set of parameters is similar to that in the previous sub-sections,
unless stated otherwise.
\begin{widetext}
We begin with investigation of the forced-damped model; since the
stable solutions of the forced-damped model generically are hyperbolic
attractors, the numeric validation of the stability analysis is easier
than in the conservative model.

\begin{figure}[H]
\begin{centering}
\includegraphics[bb=10bp 0bp 260bp 180bp,clip,width=0.7\textwidth]{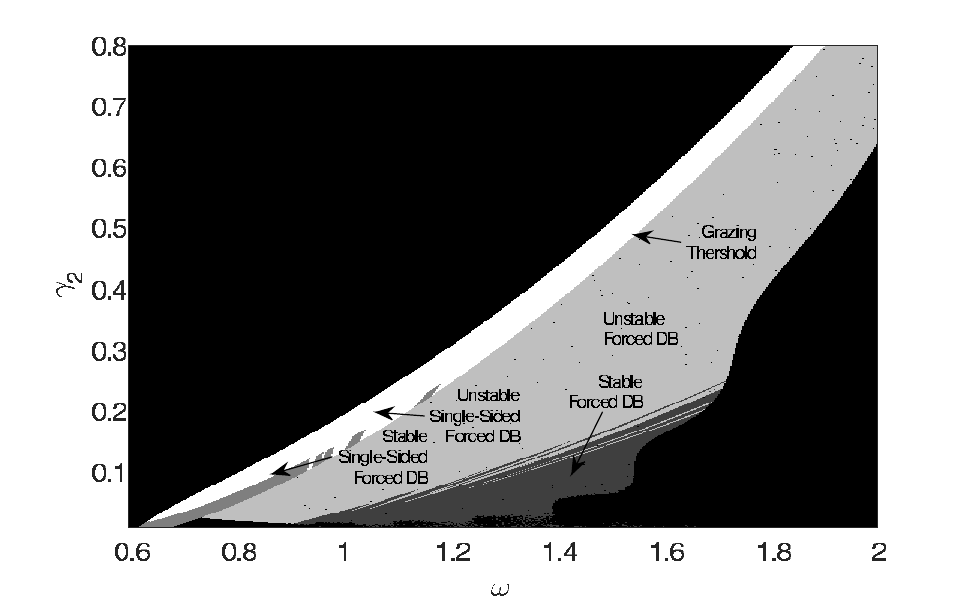}
\par\end{centering}
\caption{\label{fig:10}Existance-stability map for the forced-damped DB. The
map shows non-existing or not-physical solutions (black), stable forced
DB (dark gray), stable single-sided forced DB (gray), unstable forced
DB (light gray), and unstable single-sided forced DB (white)}
\end{figure}
\end{widetext}

Figure \ref{fig:10} shows the region of stability and existence of
the forced-damped DB in the frequency \textendash{} coupling stiffness
plane of parameters. In this map the mechanism for loss of stability
for both double- and single-sided forced DBs is Neimark-Sacker bifurcation;
however, other mechanisms for the loss of stability appear for other
sets of parameters \textendash{} the period doubling and pitchfork
bifurcations are also encountered. One can also clearly see the grazing
threshold where the forced double-sided DB surpasses the grazing point
and turns into the single-sided DB.

It is important to note that at the grazing limit, the stable solution
of the single-sided forced DB meets the unstable forced DB. The convergence
of the stability and grazing boundaries can be explained by the sudden
breaking of symmetry of the solution. While the single-sided forced
DB is symmetric with respect to the peak between two impacts, after
reaching the grazing point and with the transition to the forced DB
the second impact in the period immediately diverges from $\phi=\pi/w$,
where it occurs at the grazing point.

\begin{figure}[H]
\begin{centering}
\includegraphics[width=1\columnwidth]{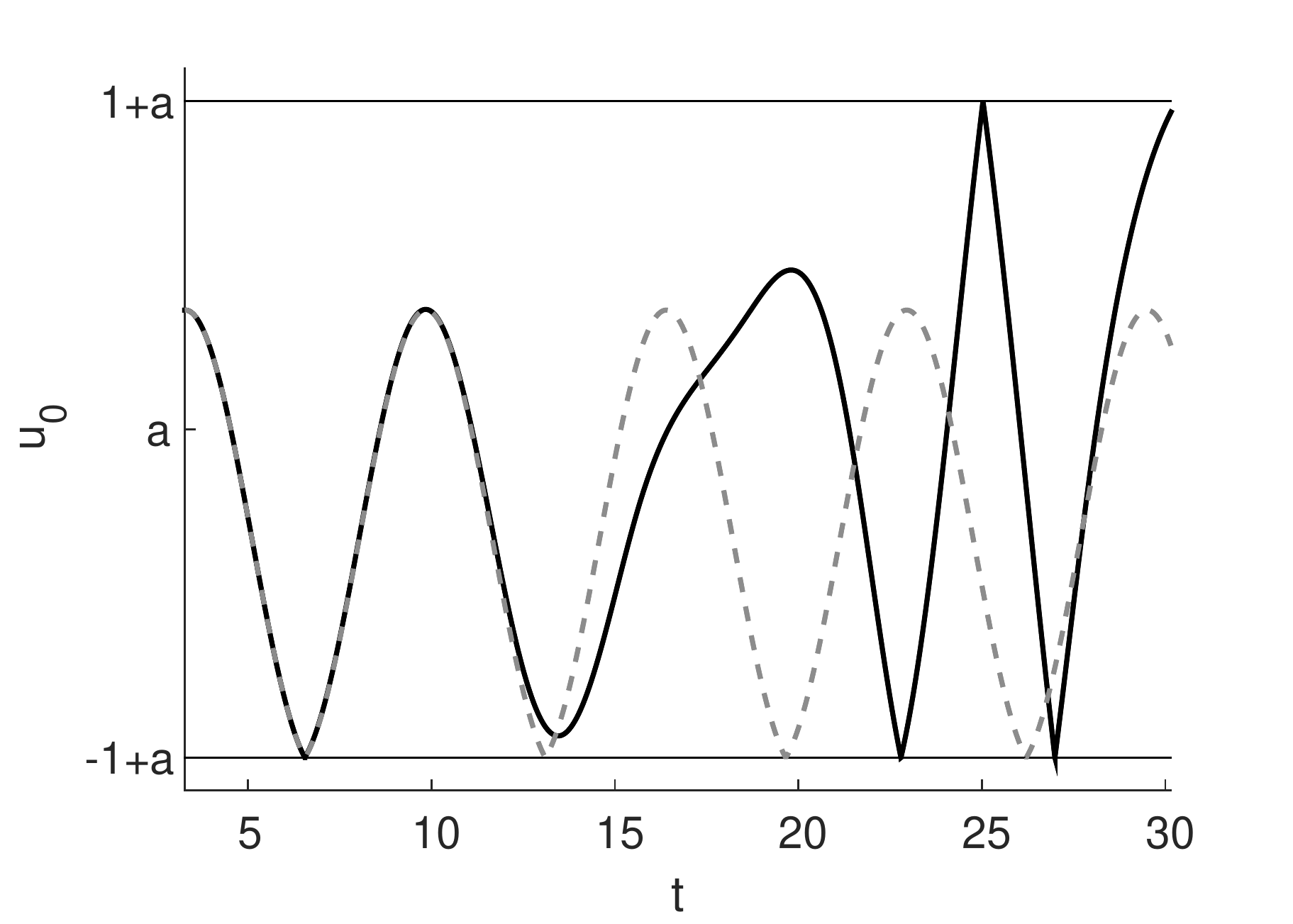}
\par\end{centering}
\caption{\label{fig:11}The displacements of the first mass (Numerical result
in black and analytic solution in dashed gray) for an unstable single-sided
forced DB via Neimark-Sacker bifurcation.}
\end{figure}

\begin{figure}[H]
\begin{centering}
\includegraphics[width=1\columnwidth]{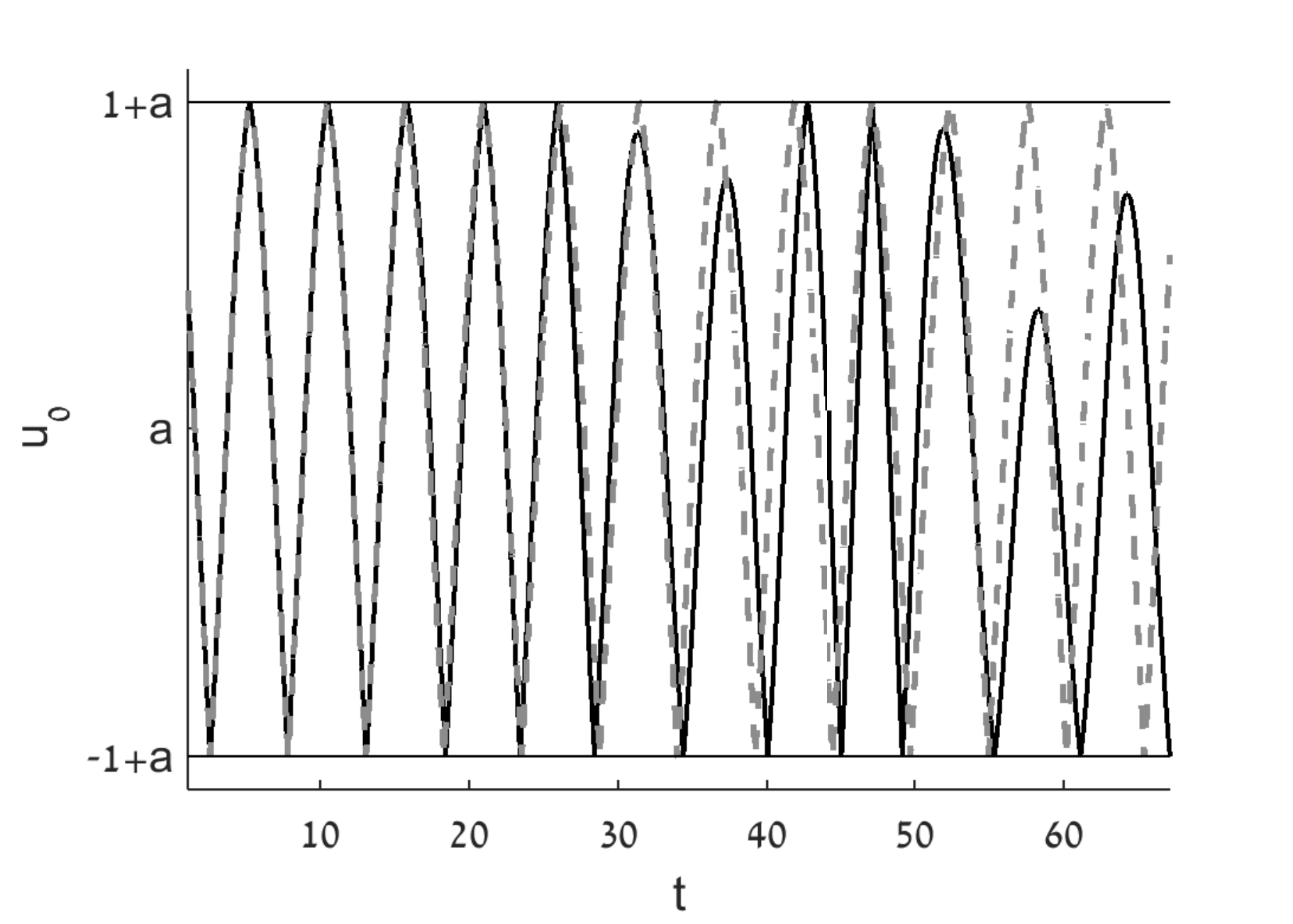}
\par\end{centering}
\caption{\label{fig:12}The displacements of the first mass (Numerical result
in black and analytic solution in dashed gray) for an unstable forced
DB via Neimark-Sacker bifurcation.}
\end{figure}

Figures \ref{fig:11} and \ref{fig:12} present examples of the loss
of stability via Neimark-Sacker bifurcation. The discrepancy between
the numerical solution and analytic prediction for the unstable solution
validates the procedure of the stability analysis presented above.

\begin{figure}[H]
\begin{centering}
\includegraphics[width=1\columnwidth]{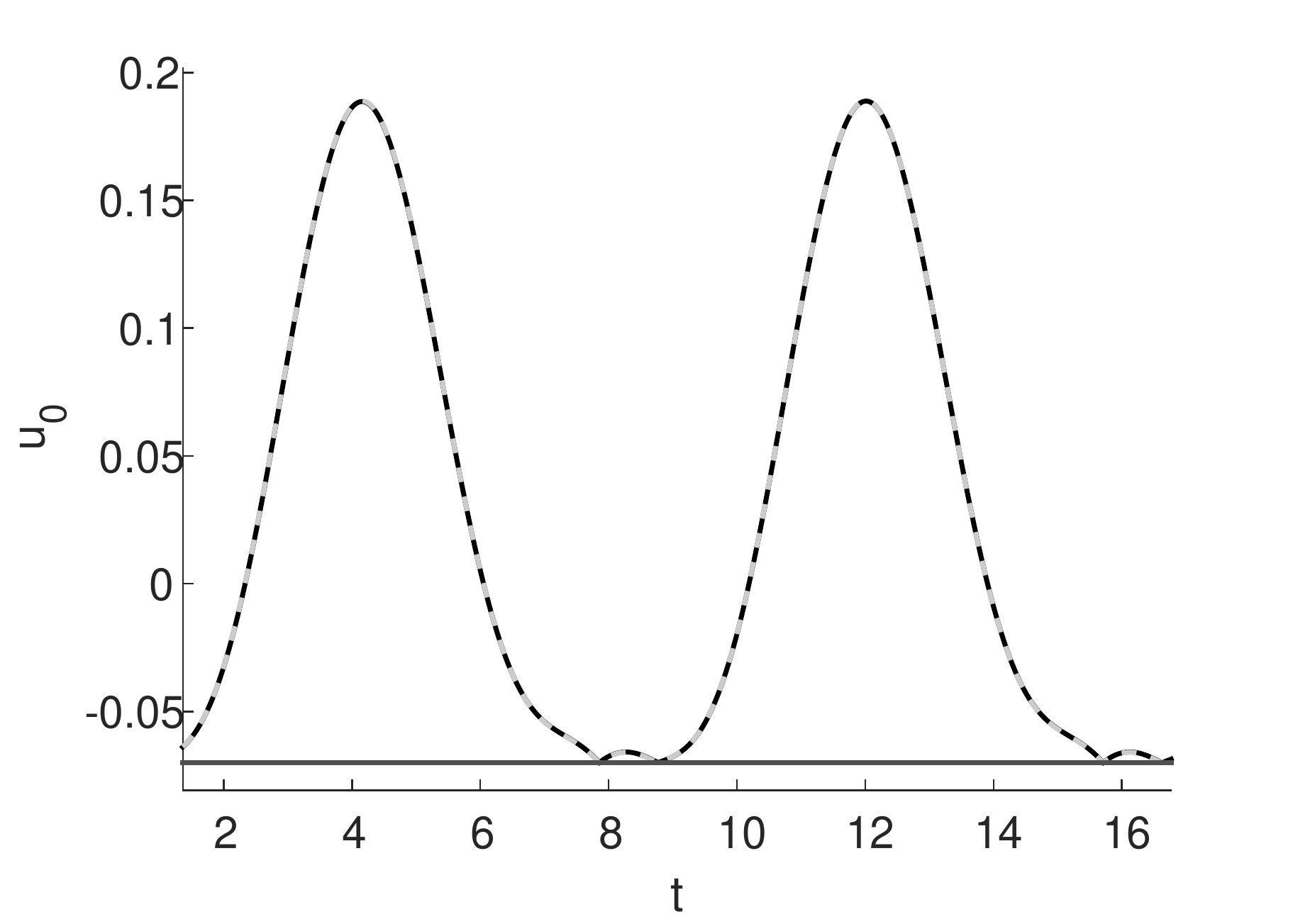}
\par\end{centering}
\caption{\label{fig:13}The displacements of the first mass (Numerical result
in black and analytic solution in dashed gray) for the forced single-sided
DB with period doubling.}
\end{figure}

As aforementioned, there are two more mechanisms for the loss of stability.
The first is the pitchfork bifurcation of the forced DB; pair of stable
asymmetric solutions is formed. These asymmetric solutions were predicted
numerically in ref. \cite{Grinberg2016} and found analytically in
this work. The last mechanism for loss of stability is through period
doubling bifurcation relevant to the single-sided forced DB. The period
doubling obtained analytically is demonstrated in fig. \ref{fig:13}.
The period doubling is clearly visible in the form to two consecutive
peaks of different heights in each period.
\begin{widetext}
~

\begin{figure}[H]
\begin{centering}
\includegraphics[width=0.45\columnwidth]{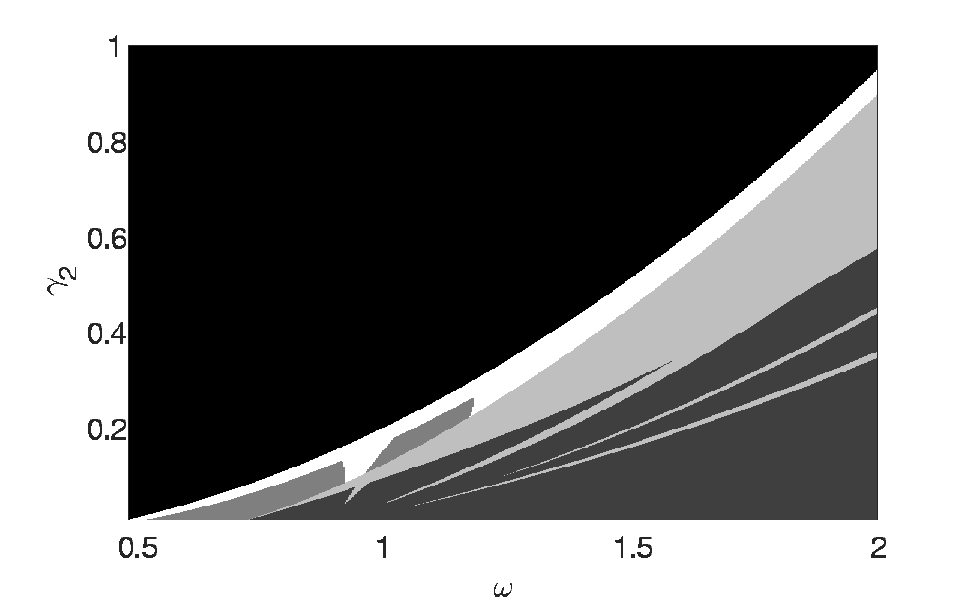}\includegraphics[width=0.45\columnwidth]{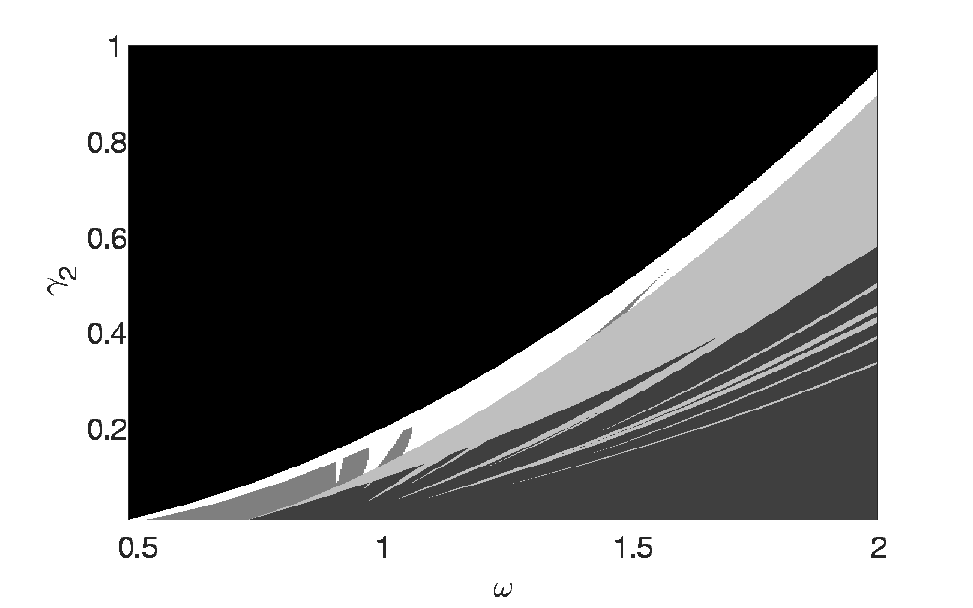}\\
 \includegraphics[width=0.45\columnwidth]{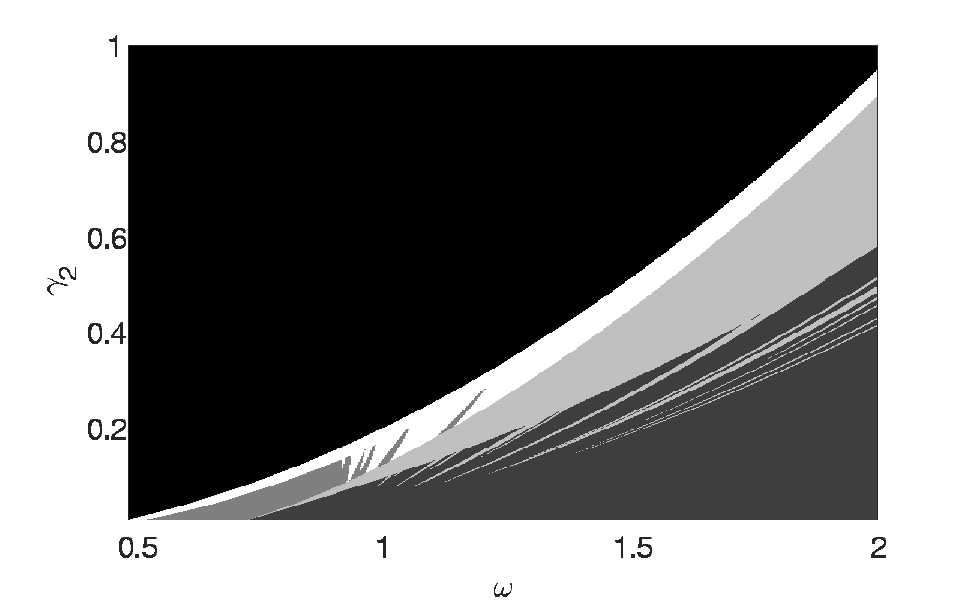}\includegraphics[width=0.45\columnwidth]{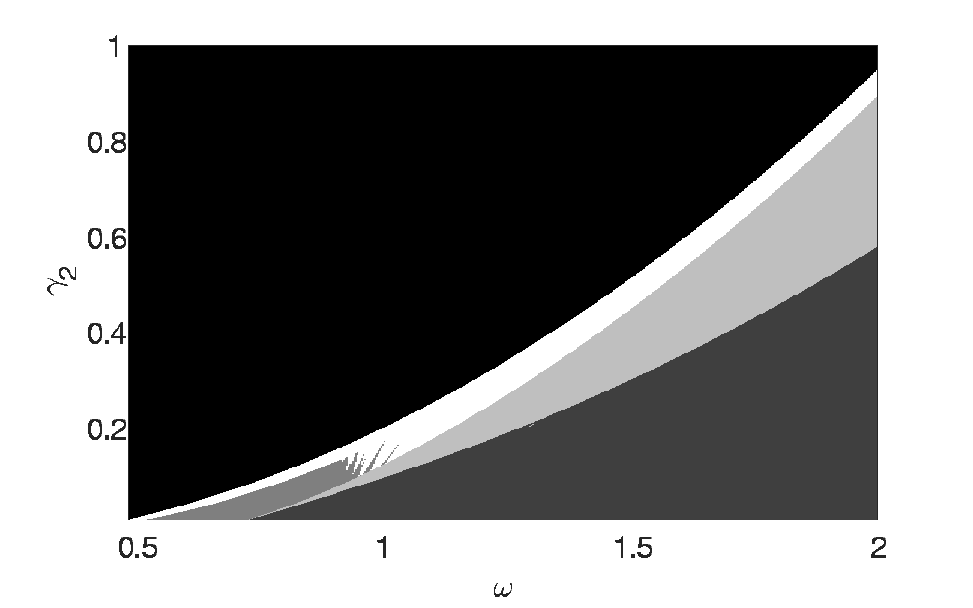}
\par\end{centering}
\caption{\label{fig:15}Existence-stability maps for a DB with $N=5$, $N=10$,
$N=20$ and $N=40$ (Top to Bottom). The map shows non-existing or
not-physical solutions (black), stable DB (dark gray), stable single-sided
DB (gray), unstable DB (light gray), and unstable single-sided DB
(white)}
\end{figure}

Figure \ref{fig:15} shows the existence-stability map for the conservative
model. There is a lot of similarity in this map to that of the forced
DB. The loss of stability here as well is via Neimark-Sacker bifurcation.
However, in the case of the conservative system it is difficult to
validate the loss of stability, since its solutions are not attractors.
Other similar features are the stripes of instability. These stripes
are related to the chain length as demonstrated in the existence-stability
maps for different chain length in fig. \ref{fig:15}, and correspond
to the spatial modes appearing in the eigenvectors of the Monodromy
matrix due to the finite size of the system, resembling the observations
in ref. \cite{Grinberg2016}. It is also interesting to note that
,with the exception of these stripes, the stability patterns are not
strongly affected by the system size.
\end{widetext}

\section{Concluding Remarks\label{sec:Concluding-Remarks}}

In this work, we derive the exact DB solutions for asymmetric vibro-impact
lattice. Contrary to the symmetric setting, two types of such solutions
exist \textendash{} the single-sided and the double-sided DBs, divided
by the grazing boundary. The asymmetric DBs can appear not only in
the intrinsically asymmetric lattice considered in the paper, but
also can result from the symmetry braking in the symmetric lattices
investigated in previous works.

Another interesting finding is the multiplicity of the stable forced
DB solutions, not observed in the symmetric lattice. It is possible
to observe more than one stable asymmetric solution for the same set
of parameters. The finite system size leads to formation of the \textquotedbl{}stripes
of instability\textquotedbl{}, which appear due to the finite set
of eigenmodes available in the finite system. This peculiarity essentially
modifies the domain of stability on the space of parameters and can
be considered as generic consequence of the finite system size.

Finally, the asymmetric setting reveal all three generic mechanisms
for loss of stability of the periodic solutions - pitchfork, Neimark-Sacker
and period doubling bifurcations. The latter was not observed in previously
studied symmetric models and an exact analytic solution for the doubled period solutions is also obtained.
\begin{acknowledgments}
The authors are very grateful to Israel Science Foundation (grant
838/13) for financial support.
\end{acknowledgments}

\bibliographystyle{plainnat}

\appendix

\section{Single-Sided Discrete Breather \textendash{} Derivations\label{sec:Single-Sided-DB-Derivation}}

The governing equations of motion are:
\begin{equation}
\ddot{u}_{0}+\gamma_{1}u_{0}+\gamma_{2}\left(2u_{0}-u_{1}-u_{N}\right)=2p\sum_{j=-\infty}^{\infty}{\delta{\left(t-\cfrac{2\pi j}{\omega}\right)}}
\end{equation}
\begin{equation}
\ddot{u}_{n}+\gamma_{1}u_{n}+\gamma_{2}\left(2u_{n}-u_{n+1}-u_{n-1}\right)=0
\end{equation}
\begin{equation}
\ddot{u}_{N}+\gamma_{1}u_{N}+\gamma_{2}\left(2u_{N}-u_{0}-u_{N-1}\right)=0
\end{equation}

The terms for the impacts can also be written in the form of generalized
Fourier series:
\begin{equation}
\ddot{u}_{0}+\gamma_{1}u_{0}+\gamma_{2}\left(2u_{0}-u_{1}-u_{N}\right)=\frac{\omega p}{\pi}\sum_{j=-\infty}^{\infty}\cos{\left(j\omega t\right)}
\end{equation}
\begin{equation}
\ddot{u}_{n}+\gamma_{1}u_{n}+\gamma_{2}\left(2u_{n}-u_{n+1}-u_{n-1}\right)=0
\end{equation}
\begin{equation}
\ddot{u}_{N}+\gamma_{1}u_{N}+\gamma_{2}\left(2u_{N}-u_{0}-u_{N-1}\right)=0
\end{equation}

This can be rewritten as follows:
\begin{equation}
\ddot{u}_{0}+\gamma_{1}u_{0}+\gamma_{2}\left(2u_{0}-u_{1}-u_{N}\right)=\frac{\omega}{\pi}p+\frac{2\omega p}{\pi}\sum_{j=1}^{\infty}{\cos{\left(j\omega t\right)}}\label{eq:7-1-1}
\end{equation}
\begin{equation}
\ddot{u}_{n}+\gamma_{1}u_{n}+\gamma_{2}\left(2u_{n}-u_{n+1}-u_{n-1}\right)=0
\end{equation}
\begin{equation}
\ddot{u}_{N}+\gamma_{1}u_{N}+\gamma_{2}\left(2u_{N}-u_{0}-u_{N-1}\right)=0\label{eq:9-1-1}
\end{equation}

Solving in a similar manner, we obtain:
\begin{equation}
u_{n}=u_{n,0}+\sum_{j=1}^{\infty}{u_{n,j}\cos{\left(j\omega t\right)}}
\end{equation}

where,
\begin{eqnarray}
u_{n,0} & = & -\frac{\omega p\left(f_{0}^{n-N-1}+f_{0}^{-n}\right)}{\pi\gamma_{2}\left(f_{0}-f_{0}^{-1}\right)\left(f_{0}^{-N-1}-1\right)}\\
u_{n,j} & = & -\frac{2\omega p\left(f_{j}^{n-N-1}+f_{j}^{-n}\right)}{\pi\gamma_{2}\left(f_{j}-f_{j}^{-1}\right)\left(f_{j}^{-N-1}-1\right)}
\end{eqnarray}

As previously, the system is closed with the help of equations, that
fix the impact at the desired location:
\begin{equation}
\begin{array}{c}
u_{0}{\left(0\right)}=-\cfrac{\omega\left(f_{0}^{-N-1}+1_{0}\right)p}{\pi\gamma_{2}\left(f_{0}-f_{0}^{-1}\right)\left(f_{0}^{-N-1}-1\right)}-\\
-\sum_{j=1}^{\infty}{\cfrac{2\omega\left(f_{j}^{-N-1}+1\right)p}{\pi\gamma_{2}\left(f_{j}-f_{j}^{-1}\right)\left(f_{j}^{-N-1}-1\right)}}=-1+a
\end{array}
\end{equation}

\begin{eqnarray}
-p\chi_{2} & = & -1+a\to p=\cfrac{1-a}{\chi_{2}}\label{eq:23-1-1}
\end{eqnarray}

\section{Single-Sided Forced-Damped Discrete Breather \textendash{} Derivations\label{sec:Single-Sided-Forced-DB-Derivations}}

The single-sided DB is also possible in the forced-damped model. The
equations of motion can be written as follows:
\begin{equation}
\begin{array}{c}
\ddot{v}_{0}+\gamma_{1}v_{0}+\gamma_{2}\left(2v_{0}-v_{1}-v_{N}\right)=\\
=F{\left(t+\psi\right)}+2p\sum_{j=-\infty}^{\infty}{\delta{\left(t-\cfrac{2\pi j}{\omega}\right)}}
\end{array}
\end{equation}
\begin{equation}
\ddot{v}_{n}+\gamma_{1}v_{n}+\gamma_{2}\left(2v_{n}-v_{n+1}-v_{n-1}\right)=F{\left(t+\psi\right)}
\end{equation}
\begin{equation}
\ddot{v}_{N}+\gamma_{1}v_{N}+\gamma_{2}\left(2v_{N}-v_{0}-v_{N-1}\right)=F{\left(t+\psi\right)}
\end{equation}
where $\psi$ is the phase external force with respect to DB's impacts.

The external force $F{\left(t\right)}$ can be removed from the equations
with the help of a simple transformation. Let $v_{n}{\left(t\right)}=u_{n}{\left(t\right)}+G{\left(t+\psi\right)}$
where $\ddot{G}{\left(t\right)}+\gamma_{1}G{\left(t\right)}=F{\left(t\right)}$.
Substitution into the above equations yields:
\begin{equation}
\ddot{u}_{0}+\gamma_{1}u_{0}+\gamma_{2}\left(2u_{0}-u_{1}-u_{N}\right)=2p\sum_{j=-\infty}^{\infty}{\delta{\left(t-\cfrac{2\pi j}{\omega}\right)}}
\end{equation}
\begin{equation}
\ddot{u}_{n}+\gamma_{1}u_{n}+\gamma_{2}\left(2u_{n}-u_{n+1}-u_{n-1}\right)=0
\end{equation}
\begin{equation}
\ddot{u}_{N}+\gamma_{1}u_{N}+\gamma_{2}\left(2u_{N}-u_{0}-u_{N-1}\right)=0
\end{equation}

Similarly, the terms for the impacts can is written in the form of
generalized Fourier series:

\begin{equation}
\ddot{u}_{0}+\gamma_{1}u_{0}+\gamma_{2}\left(2u_{0}-u_{1}-u_{N}\right)=\frac{\omega p}{\pi}\sum_{j=-\infty}^{\infty}{\cos{\left(j\omega t\right)}}
\end{equation}
\begin{equation}
\ddot{u}_{n}+\gamma_{1}u_{n}+\gamma_{2}\left(2u_{n}-u_{n+1}-u_{n-1}\right)=0
\end{equation}
\begin{equation}
\ddot{u}_{N}+\gamma_{1}u_{N}+\gamma_{2}\left(2u_{N}-u_{0}-u_{N-1}\right)=0
\end{equation}

Since the equations are identical to those of the conservative model,
the solution is similar:

\begin{equation}
u_{n}=u_{n,0}+\sum_{j=1}^{\infty}{\left(u_{n,j}\cos{\left(j\omega t\right)}\right)}
\end{equation}
where,
\begin{eqnarray}
u_{n,0} & = & -\frac{\omega p\left(f_{0}^{n-N-1}+f_{0}^{-n}\right)}{\pi\gamma_{2}\left(f_{0}-f_{0}^{-1}\right)\left(f_{0}^{-N-1}-1\right)}\\
u_{n,j} & = & -\frac{2\omega p\left(f_{j}^{n-N-1}+f_{j}^{-n}\right)}{\pi\gamma_{2}\left(f_{j}-f_{j}^{-1}\right)\left(f_{j}^{-N-1}-1\right)}
\end{eqnarray}

As in the conservative model, the solution must satisfy the impact
location equations:
\begin{equation}
v_{0}{\left(0\right)}=-p\chi_{2}+G{\left(\psi\right)}=-1+a\label{eq:48-2-2}
\end{equation}

Also, the impact law must be satisfied:
\begin{equation}
\begin{array}{c}
\dot{v}{\left(0^{+}\right)}=\dot{u}{\left(0^{+}\right)}+\dot{G}{\left(\psi\right)}=p+\dot{G}{\left(\psi\right)}=\\
=-e\left(-p+\dot{G}{\left(\psi\right)}\right)=-e\left(\dot{u}{\left(0^{-}\right)}+\dot{G}{\left(\psi\right)}\right)=-e\dot{v}{\left(0^{-}\right)}
\end{array}
\end{equation}

By further simplification, one obtains:
\begin{eqnarray}
\dot{G}{\left(\psi\right)} & = & -qp\label{eq:51-1-2}
\end{eqnarray}
where $q=\left(1-e\right)/\left(1+e\right)$.

\section{Single-Sided Forced-Damped Discrete Breather with Period Doubling
\textendash{} Derivations\label{sec:Period-Doubling-Derivations}}

We examine once more the case of a symmetric force $F{\left(t\right)}$
which satisfies $F{\left(t\right)}=F{\left(t+2\pi/\Omega\right)}$
and $F{\left(t\right)}=-F{\left(t+\pi/\Omega\right)}$, however, with
a frequency of $\Omega=2\omega$ . The solution should obey the following
set of equations:
\begin{equation}
\begin{array}{c}
\ddot{v}_{0}+\gamma_{1}v_{0}+\gamma_{2}\left(2v_{0}-v_{1}-v_{N}\right)=F{\left(t+\psi\right)}+\\
+2p_{1}\sum_{j=-\infty}^{\infty}{\delta{\left(t-\phi-\cfrac{2\pi j}{\omega}\right)}}-\\
+2p_{2}\sum_{j=-\infty}^{\infty}{\delta{\left(t-\cfrac{2\pi j}{\omega}\right)}}
\end{array}
\end{equation}
\begin{equation}
\ddot{v}_{n}+\gamma_{1}v_{n}+\gamma_{2}\left(2v_{n}-v_{n+1}-v_{n-1}\right)=F{\left(t+\psi\right)}
\end{equation}
\begin{equation}
\ddot{v}_{N}+\gamma_{1}v_{N}+\gamma_{2}\left(2v_{N}-v_{0}-v_{N-1}\right)=F{\left(t+\psi\right)}
\end{equation}
where $\psi$ is the phase of the external force with respect to DB's
impacts.

The external force $F{\left(t\right)}$ can be removed from the equations
with the help of a simple transformation. Let $v_{n}{\left(t\right)}=u_{n}{\left(t\right)}+G{\left(t+\psi\right)}$
where $\ddot{G}{\left(t\right)}+\gamma_{1}G{\left(t\right)}=F{\left(t\right)}$.
Substitution into the above equations yields:
\begin{equation}
\begin{array}{c}
\ddot{u}_{0}+\gamma_{1}u_{0}+\gamma_{2}\left(2u_{0}-u_{1}-u_{N}\right)=\\
=2p_{1}\sum_{j=-\infty}^{\infty}{\delta{\left(t-\phi-\cfrac{2\pi j}{\omega}\right)}}-\\
+2p_{2}\sum_{j=-\infty}^{\infty}{\delta{\left(t-\cfrac{2\pi j}{\omega}\right)}}
\end{array}
\end{equation}
\begin{equation}
\ddot{u}_{n}+\gamma_{1}u_{n}+\gamma_{2}\left(2u_{n}-u_{n+1}-u_{n-1}\right)=0
\end{equation}
\begin{equation}
\ddot{u}_{N}+\gamma_{1}u_{N}+\gamma_{2}\left(2u_{N}-u_{0}-u_{N-1}\right)=0
\end{equation}

Similarly, the terms for the impacts is written in the form of generalized
Fourier series:

\begin{equation}
\begin{array}{c}
\ddot{u}_{0}+\gamma_{1}u_{0}+\gamma_{2}\left(2u_{0}-u_{1}-u_{N}\right)=\\
=\frac{\omega}{\pi}\sum_{j=-\infty}^{\infty}{\left(p_{1}\cos{\left(j\omega\left(t-\phi\right)\right)}+p_{2}\cos{\left(j\omega t\right)}\right)}
\end{array}
\end{equation}
\begin{equation}
\ddot{u}_{n}+\gamma_{1}u_{n}+\gamma_{2}\left(2u_{n}-u_{n+1}-u_{n-1}\right)=0
\end{equation}
\begin{equation}
\ddot{u}_{N}+\gamma_{1}u_{N}+\gamma_{2}\left(2u_{N}-u_{0}-u_{N-1}\right)=0
\end{equation}

The equations are identical to those of the conservative model. Hence,
the solution is similar:

\begin{equation}
u_{n}=u_{n,0}+\sum_{j=1}^{\infty}{\left(u_{n,j,1}\cos{\left(j\omega\left(t-\phi\right)\right)}+u_{n,j,2}\cos{\left(j\omega t\right)}\right)}
\end{equation}

where
\begin{eqnarray}
u_{n,0} & = & -\frac{\omega\left(p_{2}+p_{1}\right)\left(f_{0}^{n-N-1}+f_{0}^{-n}\right)}{\pi\gamma_{2}\left(f_{0}-f_{0}^{-1}\right)\left(f_{0}^{-N-1}-1\right)}\\
u_{n,j,1} & = & -\frac{2\omega p_{1}\left(f_{j}^{n-N-1}+f_{j}^{-n}\right)}{\pi\gamma_{2}\left(f_{j}-f_{j}^{-1}\right)\left(f_{j}^{-N-1}-1\right)}\\
u_{n,j,2} & =- & \frac{2\omega p_{2}\left(f_{j}^{n-N-1}+f_{j}^{-n}\right)}{\pi\gamma_{2}\left(f_{j}-f_{j}^{-1}\right)\left(f_{j}^{-N-1}-1\right)}
\end{eqnarray}

As in the conservative setting, the solution must satisfy the impact
location equations:
\begin{equation}
v_{0}{\left(0\right)}=-p_{1}\chi_{1}{\left(\phi\right)}-p_{2}\chi_{2}+G{\left(\psi\right)}=-1+a\label{eq:47-1}
\end{equation}
\begin{equation}
v_{0}{\left(\phi\right)}=-p_{1}\chi_{2}-p_{2}\chi_{1}{\left(\phi\right)}+G{\left(\psi+\phi\right)}=-1+a\label{eq:48-1}
\end{equation}

Also, the impact law must be satisfied:
\begin{equation}
\begin{array}{c}
\dot{v}_{0}{\left(0^{+}\right)}=\dot{u}_{0}{\left(0^{+}\right)}+\dot{G}{\left(\psi\right)}=\\
=-e\left(\dot{u}_{0}{\left(0^{-}\right)}+\dot{G}{\left(\psi\right)}\right)=-e\dot{v}_{0}{\left(0^{-}\right)}
\end{array}
\end{equation}

\begin{equation}
\dot{u}_{0}{\left(0^{+}\right)}+e\dot{u}_{0}{\left(0^{-}\right)}=-\dot{G}{\left(\psi\right)}\left(1+e\right)\label{eq:71-1}
\end{equation}

The generalized Fourier series converges to the average of the velocities
on both sides of the discontinuity:

\begin{equation}
\cfrac{\dot{u}_{0}{\left(0^{+}\right)}+\dot{u}_{0}{\left(0^{-}\right)}}{2}=-p_{1}\chi_{3}
\end{equation}

Conservation of momentum during the impact yields:
\begin{equation}
\cfrac{\dot{u}_{0}{\left(0^{+}\right)}-\dot{u}_{0}{\left(0^{-}\right)}}{2}=p_{2}
\end{equation}

From these equations we extract terms for the velocities:
\begin{equation}
\dot{u}_{0}{\left(0^{+}\right)}=p_{2}-p_{1}\chi_{3}
\end{equation}
\begin{equation}
\dot{u}_{0}{\left(0^{-}\right)}=-p_{2}-p_{1}\chi_{3}
\end{equation}

Note that the change in energy for the reduced un-forced system during
the impact is $\Delta E_{1}=-2p_{1}p_{2}\chi_{3}$.

Plugging into eq. (\ref{eq:71-1}), one obtains:
\begin{equation}
\dot{G}{\left(\psi\right)}=p_{1}\chi_{3}-qp_{2}\label{eq:75-1}
\end{equation}
where $q=\left(1-e\right)/\left(1+e\right)$.

Similarly, it is possible to perform the same procedure for the second
impact:

\begin{equation}
\dot{u}_{0}{\left(\phi^{+}\right)}+e\dot{u}{\left(\phi^{-}\right)}=-\dot{G}{\left(\phi+\psi\right)}\left(1+e\right)\label{eq:78-1}
\end{equation}

The generalized Fourier series converges to the average of the velocities
on both sides of the discontinuity:

\begin{equation}
\cfrac{\dot{u}{\left(\phi^{+}\right)}+\dot{u}{\left(\phi^{-}\right)}}{2}=p_{2}\chi_{3}
\end{equation}

Conservation of momentum during the impact yields:
\begin{equation}
\cfrac{\dot{u}{\left(\phi^{+}\right)}-\dot{u}{\left(\phi^{-}\right)}}{2}=p_{1}
\end{equation}

From these equations we extract the terms for the velocities:
\begin{equation}
\dot{u}{\left(\phi^{+}\right)}=p_{1}+p_{2}\chi_{3}
\end{equation}
\begin{equation}
\dot{u}{\left(\phi^{-}\right)}=-p_{1}+p_{2}\chi_{3}
\end{equation}

Note that the energy gain in this impact is $\Delta E_{2}=2p_{1}p_{2}\chi_{3}$;
hence the energy of the reduced un-forced system is conserved throughout
the period as expected.

Plugging into eq. (\ref{eq:78-1}):
\begin{equation}
\dot{G}{\left(\phi+\psi\right)}=-p_{2}\chi_{3}-qp_{1}\label{eq:81-1}
\end{equation}

\subsubsection{Harmonic Excitation}

In order to solve the equations we need to choose the forcing function,
that satisfies the symmetry conditions. Let us choose $F{\left(t\right)}=A\cos{\left(2\omega t\right)}$.
Solving the ODE, we obtain:
\begin{equation}
G{\left(t\right)}=\tilde{A}\cos{\left(2\omega t\right)}
\end{equation}
where,
\begin{equation}
\tilde{A}=\cfrac{A}{\gamma_{1}-\left(2\omega\right)^{2}}
\end{equation}

Plugging the solution into eq. (\ref{eq:47-1}), (\ref{eq:48-1}),
(\ref{eq:75-1}) and (\ref{eq:81-1}), one obtains the following expressions:
\begin{equation}
-p_{1}\chi_{1}{\left(\phi\right)}-p_{2}\chi_{2}+\tilde{A}\cos{\left(2\omega\psi\right)}=-1+a
\end{equation}
\begin{equation}
-p_{1}\chi_{2}-p_{2}\chi_{1}{\left(\phi\right)}+\tilde{A}\cos{\left(2\omega\left(\psi+\phi\right)\right)}=-1+a
\end{equation}

\begin{equation}
-2\tilde{A}\omega\sin{\left(2\omega\left(\psi+\phi\right)\right)}=-p_{2}\chi_{3}-qp_{1}
\end{equation}
\begin{equation}
-2\tilde{A}\omega\sin{\left(2\omega\psi\right)}=p_{1}\chi_{3}-qp_{2}
\end{equation}

Note that for $\phi=\pi/\omega$ we obtain this set of equation is
reduced to two independent equations and we get the regular single-sided
forced DB solution.

To find the exact solution explicitly, we assume that $\phi$ is known
and the barrier asymmetry $a$ is the unknown. Solution of the above
set of equations under this assumption yields:
\begin{equation}
\psi=\cfrac{\pm\frac{\pi}{2}+\alpha}{2\omega}
\end{equation}
\begin{equation}
p_{1}=\cfrac{2\tilde{A}\omega\left(q\sin{\left(2\omega\left(\psi+\phi\right)\right)}-\chi_{3}\sin{\left(2\omega\psi\right)}\right)}{q^{2}+\chi_{3}^{2}}
\end{equation}
\begin{equation}
p2=-\cfrac{2\tilde{A}\omega\left(\chi_{3}\sin{\left(2\omega\left(\psi+\phi\right)\right)}+q\sin{\left(2\omega\psi\right)}\right)}{q^{2}+\chi_{3}^{2}}
\end{equation}
\begin{equation}
a=-p_{1}\chi_{1}{\left(\phi\right)}-p_{2}\chi_{2}+\tilde{A}\cos{\left(2\omega\psi\right)}+1
\end{equation}

\begin{widetext}
where,
\begin{equation}
\sigma=\sqrt{\begin{array}{c}
8\omega^{2}\chi_{3}^{2}\left(\chi_{1}-\chi_{2}\right)^{2}\left(1+\cos{\left(2\omega\phi\right)}\right)+8\omega\chi_{3}\left(q^{2}+\chi_{3}^{2}\right)\left(\chi_{1}-\chi_{2}\right)\sin{\left(2\omega\phi\right)}+\\
+2\left(\chi_{3}^{4}-q^{4}+q^{2}\left(4\omega^{2}\left(\chi_{1}-\chi_{2}\right)^{2}+2\chi^{3}\right)\right)\left(1-\cos{\left(2\omega\phi\right)}\right)
\end{array}}
\end{equation}
\begin{equation}
\alpha=\pm\arccos{\left(\cfrac{\left(q^{2}+\chi_{3}^{2}\right)\left(1-\cos{\left(2\omega\phi\right)}\right)-\left(q-\chi_{3}\right)\left(\chi_{1}-\chi_{2}\right)\omega\sin{\left(2\omega\phi\right)}}{\sigma}\right)}
\end{equation}
\end{widetext}

\end{document}